%% file: main.tex
\documentclass[a4paper,onecolumn,11pt,noarxiv]{quantumarticle}

\usepackage[utf8]{inputenc}
\usepackage[english]{babel}
\usepackage[T1]{fontenc}

\usepackage{amsmath,amsthm,amsfonts,amssymb,dsfont,graphicx,mathtools,physics,thmtools,thm-restate}
\usepackage{xprintlen}

\usepackage{complexity}
\usepackage{enumitem}
\usepackage{fullpage}

\usepackage{comment}
\usepackage{algorithm}  
\usepackage{algpseudocode}
\algrenewcommand\algorithmicrequire{\textbf{Input:}}
\algrenewcommand\algorithmicensure{\textbf{Output:}}

\usepackage{xcolor}

\usepackage{xurl}
\usepackage{hyperref}

\hypersetup{
    breaklinks=true,
  colorlinks=true,
  hypertexnames=false,
  linktocpage,
  colorlinks=true, 
  urlcolor=magenta!90!black,    
  linkcolor=blue!60!black, 
  citecolor=black!60 
}
\usepackage[capitalize, compress]{cleveref}
\usepackage[
  backend=biber,
  citestyle=alphabetic,
  bibstyle=alphabetic,
  minalphanames=3,maxalphanames=4,
  maxnames=10,minnames=1,
  maxcitenames=3,mincitenames=1,
  eprint=true,
  date=year,
  doi=false,
  url=false,
  isbn=false,
  ]{biblatex}
\AtEveryBibitem{
  \clearfield{urldate}
  \clearfield{primaryclass}
}

\input{mymath}

\newcommand{\simons}{%
Simons Institute for the Theory of Computing, University of California at Berkeley}
\newcommand{\ucb}{%
University of California at Berkeley}
\newcommand{\ethz}{%
ETH Z\"urich}

\title{Code Swendsen-Wang Dynamics}

\author{Dominik Hangleiter}
\affiliation{\simons}
\affiliation{\ethz}
\author{Nathan Ju}
\affiliation{\ucb}
\author{Umesh Vazirani}
\affiliation{\ucb}

\date{}

\begin{document}
\maketitle
\begin{abstract}

Recent advances in quantum Gibbs sampling leave open the central question of rapid mixing near and below phase transitions. This challenge is especially relevant for code Hamiltonians whose Gibbs states capture phenomena such as the thermal stability of quantum topological order. In this work, we formulate a new Markov chain, \emph{Code Swendsen-Wang dynamics}, which uses global updates to prepare the Gibbs states of arbitrary code Hamiltonians. We establish Code Swendsen-Wang dynamics as the right generalization of Swendsen-Wang dynamics for the Ising model to quantum and classical code Hamiltonians: it mixes rapidly for all previously known code Hamiltonians with efficient Gibbs samplers, resolves the central open case of the 4D toric code, and meets fundamental barriers exactly at first-order phase transitions.

\end{abstract}

\setcounter{tocdepth}{2}
\tableofcontents


\section{Introduction}
\label{sec:introduction}

Recently, there have been tremendous advances in algorithms for quantum Gibbs sampling---the task of preparing quantum Gibbs states. The main open problem is establishing rapid mixing of these chains, thereby establishing efficiency of the Markov chain.

At high temperatures, recently proposed algorithms based on (quasi-)local dynamics do mix rapidly to the Gibbs state~\cite{kastoryano_quantum_2016,brandao_finite_2019,yin_polynomial-time_2023,bakshi_high-temperature_2024,rouze_efficient_2025}. At low temperatures, these dynamics also mix rapidly in the absence of phase transitions \cite{alicki_thermalization_2009,bergamaschi_quantum_2025}. However, the most interesting regime for physics and the most difficult for algorithms is near and below critical points, where macroscopic quantities drastically change and local dynamics mix only after exponentially long times. This slowdown is caused by thermally stable phases, or modes of the Gibbs state separated by energy barriers, that local dynamics cannot traverse except with exponentially small probability~\cite{alicki_thermalization_2009,gamarnik_slow_2024,rakovszky_bottlenecks_2024}.

The same exponential slowdown plagues local chains for classical systems, such as the Ising model, which serve as minimal models capturing essential phenomenology of magnetism and (continuous) phase transitions. 
It was not until the introduction of Markov chains employing global updates by \textcite{swendsen_nonuniversal_1987}  that computer simulations could access these systems near their critical points. 
Just as the Ising model captures the essence of classical phase transitions, quantum code Hamiltonians are minimal models of quantum topological phases and their phase transitions \cite{kitaev_fault-tolerant_2003,placke_topological_2024}. 
Accessing these systems near and below critical points requires new Markov chains for code Hamiltonians that overcome the obstacles posed by energy barriers near and below phase transitions.

Inspired by the success of the Swendsen-Wang chain for sampling classical spin systems, quantum Markov chains employing global updates have recently been proposed for quantum systems \cite{ding_polynomial-time_2024, chen2025quantumreplicaexchange,paez-velasco_efficient_2025}, but none have been successful at showing rapid mixing for systems with extensive energy barriers. A major problem appears to be that the classical SW chain and its generalizations are tailored to systems with pairwise interactions, making it inapplicable to the higher-order interactions of code Hamiltonians. 

In this paper, we introduce a global-update Markov chain for preparing and simulating Gibbs states of commuting Hamiltonians, the \emph{Code Swendsen-Wang (CSW)} chain. 
The CSW chain prepares the Gibbs states of arbitrary code Hamiltonians and traverses energy landscapes with extensive barriers by employing global updates.
In particular, it is the first Markov chain that  mixes rapidly for the 4D toric code, the canonical toy model of finite-temperature topological order and the central example of a system for which local dynamics fails to mix, from any initial configuration at any temperature. 
More generally, we prove that it mixes rapidly for Hamiltonians with an approximate ``graphic'' or ``cographic'' representation, which includes all code Hamiltonians whose Gibbs states have been known to have rapidly mixing Markov chains. 
We also delineate rigorously the mixing-time properties of the CSW chain, showing that it can face exponential bottlenecks at first-order phase-transition points.

\subsection{The Code Swendsen-Wang chain}
\label{sub:the_code_swendsen_wang_chain}

The key conceptual contribution of this paper is to formulate a natural generalization, the CSW dynamics, of the SW dynamics to quantum and classical error-correcting codes.
We first outline the chain for classical codes and detail below how it is applied to prepare quantum Gibbs states. 
The objective of this chain is to generate noisy codewords\footnote{Here we assume a binary alphabet, but the generalization to larger alphabets is straightforward.} $\sigma \in \{\pm 1\}^n$ from the Gibbs distribution
\begin{equation}\label{eq:classical gibbs}
  \pi(\sigma) \propto e^{- \beta H(\sigma)}
\end{equation}
of a classical code defined by a set of parity checks with associated energy function
\begin{equation}\label{eq:cch}
  H(\sigma) = - \sum_{A \, \in \, \text{checks}} \left(\prod_{i \in A} \sigma_i \right).
\end{equation}
The Code SW dynamics achieves this by iterating updates of the noisy codeword configurations and cluster configurations.
\begin{itemize}
  \item \emph{Cluster formation:} Given a noisy codeword $\sigma$, let $E(\sigma)$ be the checks satisfied by $\sigma$. Remove checks from $E(\sigma)$ iid.\ with probability $e^{-2 \beta}$, resulting in $S \subset E(\sigma)$. 
  \item \emph{Cluster update:} Sample a new noisy codeword $\sigma'$ by sampling a random codeword from the code defined by the checks $S$, i.e., sample $\sigma'$ uniformly such that $\prod_{i \in A} \sigma'_i = 1$ for all $A \in S$. 
\end{itemize}
We show that the stationary distribution of this chain is indeed given by the Gibbs distribution~\eqref{eq:classical gibbs}. 

In fact, the SW dynamics for the Ising model is a special case of the Code SW dynamics where the interactions are restricted to be pairwise. In \cref{eq:cch}, let $H(\sigma)$ be the Ising Hamiltonian, where each check is a pairwise check between adjacent vertices in the interaction graph.
The first step of the SW chain first samples a random subgraph of $E(\sigma)$, the set of edges along which spins align. This is an identical step to CSW's cluster formation step. 
In the second step, each connected component, termed a ``cluster,'' of the resulting subgraph is then assigned a new spin in $\{+1, -1\}$ independently and uniformly, yielding a new spin configuration. By identifying this space of spin configurations as a linear subspace, this step becomes identical to CSW's cluster update step.

The CSW chain can be applied to both generate quantum samples from and classically simulate the Gibbs states of arbitrary stabilizer codes. We outline the essential idea for CSS codes. 
The goal of the algorithm is then to prepare the Gibbs state $\rho_\beta \propto e^{-\beta H}$ for the Hamiltonian
\begin{align*}
H = -\sum_{A \in \text{X checks}} X_A - \sum_{A \in \text{Z checks}} Z_A,
\end{align*}
where $X_A$ denotes the Pauli-$X$ operator supported on the qubits in $A$, and likewise for $Z_A$. 
Observe that by measuring the stabilizer operators, we project the system into an eigenstate of $H$, by which point our task reduces to sampling an eigenstate whose energy is distributed according to the correct (Gibbs) distribution. 
To achieve this, in each step of the quantum Code SW chain, we measure $X$ and $Z$ stabilizers and apply complementary $Z$ and $X$ errors drawn from CSW chains for the $X$ and $Z$ code, respectively. 
Because the two sectors evolve independently through the course of the quantum Markov chain, the mixing time of the quantum chain is determined by the slower of the two classical chains. 
This chain can be simulated classically using the stabilizer formalism \cite{gottesman_stabilizer_1997}. 

\subsection{Mixing-time results}
\label{sub:main_results}

Our main technical contributions concern a broad characterization of the mixing properties of Code SW dynamics.
First, we show rapid mixing for a large class of codes at any temperature. 
In particular, our results imply the first algorithm for the rapid preparation of the Gibbs state of the 4D toric code at any temperature, and starting from an arbitrary initial state.
Second, we show that Code SW dynamics can suffer from torpid mixing for a $p$-spin model at a \emph{first order} phase transition, a similar bottleneck faced by SW dynamics for the $q$-state Potts model with $q \ge 3$. 

To state our rapid mixing results, we introduce the notion of an approximately \emph{graphic} and \emph{cographic} parity check matrix $h \in \bin^{c \times n}$ of a classical linear code, a  notion that is borrowed from the literature on binary matroids \cite{oxley_matroid_2011}. 
We say that $h$ is graphic if the linear dependencies of its rows (its matroid) are captured by a graph, i.e., if $\ker(h^T) = \ker(g^T)$ for the edge-vertex incidence matrix $g$ of a graph. 
It is cographic if there is a graphic generator matrix for the \emph{syndrome space} of the code, i.e., the possible patterns of violated checks given by $\col(h)$, its dual matroid.  
We can relax this notion to parity checks that are \emph{close} to being graphic or cographic. 
We say that $h$ is $\Delta$-graphic if \emph{most} (all but $\Delta$ many) of its linear dependencies are captured by a graph, i.e., if $\ker(h^T)$ is a subspace of $\ker(g^T)$ for graphic $g$ with codimension~$\Delta$. 
\begin{theorem}[Rapid mixing for $\Delta$-graphic or $\Delta$-cographic codes]
\label{cor:general rapid mixing statement intro}
  Given a parity check matrix~$h$, the Code SW algorithm for the Gibbs distribution of~$h$ mixes in time $2^{\Delta} \cdot \poly(n)$ at any temperature if~$h$ is $\Delta$-graphic or $\Delta$-cographic.
\end{theorem}
\noindent
Due to a classic algorithm by \textcite{tutte_algorithm_1960}, there is an efficient algorithm to decide whether a parity check matrix is $\Delta$-(co)graphic.\footnote{Tutte's algorithm decides in $\poly(n)$ time whether a matroid (parity check matrix) is graphic \cite[]
[Prop.\ 9.4.23]{oxley_matroid_2011}. Thus, we can check if a parity check matrix is $\Delta$-graphic in $\poly(n^{\Delta}, n)$ time by checking graphicness for every $\binom{n}{\Delta}$ subspace.}

Simple examples of codes to which the theorem applies are $0$-graphic codes (i.e., $\Delta =0$). Good LDPC codes \cite{hong_quantum_2025,placke_topological_2024} and the surface code (which have no linear dependencies) correspond to a line graph.
Meanwhile, the 2D toric code (with one global linear dependency) corresponds to a simple cycle. 
Larger values of $\Delta$ (up to $O(\log n)$) yield a broader class of code Hamiltonians to which the theorem applies, and in fact encompass all code Hamiltonians for which we know of efficient Gibbs state preparation algorithms~\cite{alicki_thermal_2010, ding_polynomial-time_2024, schmidhuber_hamiltonian_2025, paez-velasco_efficient_2025, shum2025gibbssampling}. 
In contrast to these algorithms, the CSW dynamics is also conceptually simpler.

The most important nontrivial examples to which our theorem applies are  codes that arise from any chain complex on a two-dimensional surface, i.e. parity check $h$ which has the property that $g^T h=0$ for the edge-vertex incidence matrix $g$ of a graph. This is the case in particular for the $X$ and $Z$ checks of the 4D toric code, where $\Delta=4$, implying our central result. 
\begin{corollary}
  The code SW algorithm for the 4D toric code mixes rapidly at any temperature. 
\end{corollary}
The $\Delta$-cographic property of the $D$-dimensional toric code can be understood as follows: the syndrome space of the toric code is generated by all trivial cycles of the $D$-dimensional lattice, whereas the space of all cycles of the lattice is its dual. 
The topologically nontrivial cycles are therefore precisely the missing generators to make the syndrome space cographic. For a $D$-dimensional square lattice there are exactly $D$ such cycles, yielding $\Delta = 4$ for the 4D toric code. 

Intuitively, $\Delta$-(co)graphicness constrains the mixing time exponentially for the following reason: We know how to sample from all cycles with the correct distribution \cite{jerrum_polynomial-time_1990}, but to use this algorithm to sample from the correct Gibbs distribution, we now need to postselect on valid cycles, which incurs a cost of $2^\Delta$.

We also note that the Code SW dynamics naturally allow us to sample from Ising models with local fields, where corresponding Ising SW dynamics were only recently shown to mix rapidly~\cite{feng_swendsen-wang_2023}.  
Our proof of rapid mixing directly extends to this case. 
An interesting example a code for which this implies rapid mixing is the $Z$ part of the  2D toric code with local fields on the qubits corresponding to the vertical edges of the 2D lattice. 
Altogether, using the correspondence between the quantum and classical chains, and the  series of classical Markov chains just described, we can prove that our Gibbs sampling algorithm is efficient at all temperatures for code Hamiltonians with positive coefficients whose classical checks are close to being graphic or cographic with additional local fields.

Our second result studies whether the Code SW algorithm faces any obstacles to rapid mixing. This is an interesting question, since the SW algorithm for the Ising model always mixes rapidly, while it is known that the SW algorithm for the $q$-state Potts model with $q \ge 3$ faces exponential obstacles~\cite{gore_swendsenwang_1999,borgs_tight_2012,galanis_swendsen-wang_2019,gheissari_exponentially_2018}.
Indeed, we show that the same phenomenon occurs for Code SW dynamics applied to \emph{3-spin Curie-Weiss} model with energy function
\begin{equation}
\label{eq:curie weiss intro}
H(\sigma) = -\sum_{i<j<k}  \sigma_i \sigma_j \sigma_k.
\end{equation}
\begin{theorem}[Torpid mixing at first-order phase transitions]
There exists an inverse temperature $\beta^* > 0 $  at which the Code SW algorithm mixes in time $\exp(\Omega(n))$ for the 3-spin Curie-Weiss model.
\end{theorem}
This shows that the Code SW dynamics face obstacles near first order phase transition points.

\subsection{Phenomenology of the mixing behaviour}

Here we discuss the phenomenology of the aforementioned mixing properties of the CSW dynamics and relate it to the type of phase transition (first vs. second) that the system undergoes.

Recall that the Gibbs state at a given temperature is essentially determined by the free energy as a function of a natural order parameter; for the models we consider, this order parameter is the magnetization (sum of the spin values). 
The free energy captures the tradeoff between the energy contribution and the number of configurations at a given magnetization. 
The idea of a \emph{phase} captures qualitatively distinct patterns of free-energy minima at different temperatures. 

\begin{figure}
    \centering
    \includegraphics{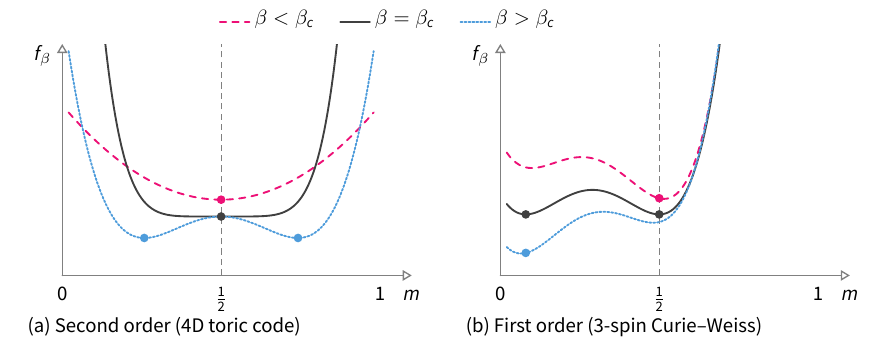}
    \caption{Illustration of the free energy landscape of second- and first-order phase transitions as a function of the magnetization order parameter $m$ above (red dashed), at (black) and below (blue dotted) the critical inverse temperature $\beta_c$. 
    (a) At a second-order transition (such as for the 2D Ising model or 4D toric code) a disordered global minimum at $m=1/2$ flattens out to a zero-curvature minimum and then splits into distinct minima at finite magnetization that are separated by a free-energy barrier below the critical temperature. Throughout the phase diagram, the free energy is symmetric around the $\mb Z_2$ symmetry exchanging $0$ and $1$. 
    (b) At a first-order transition (such as for the 3-spin Curie-Weiss model) the free energy has distinct local minima throughout the temperature range. Above the critical temperature, the disordered $m=1/2$ minimum is the global minimum; at the critical point both minima are global minima  separated by an energy barrier; at low temperatures, the only global minimum is one with finite magnetization. 
    Importantly, the distinct minima at the critical point are not related by a symmetry of the model. 
    }
    \label{fig:phase transitions}
\end{figure}

In the 4D toric code, at high temperatures ($\beta$ below the critical inverse temperature), there is a ``disordered'' phase where the spins are random and have a unique minimum of the free energy at $0$ magnetization; at low temperatures ($\beta$ above the critical inverse temperature), there is an ``ordered'' phase with two distinct minima at nonzero magnetization. The model undergoes a so-called second order transition as the temperature is lowered from disordered into ordered phase, meaning that the minimum flattens out to zero curvature at the critical point and then splits into distinct minima below it, see \cref{fig:phase transitions}.

In the 3-spin Curie Weiss model, there is also a disordered phase at high temperature, while at low temperature, there is a unique minimum at finite positive magnetization, constituting the ordered phase. In contrast to the 4D toric code, the $3$-spin Curie Weiss model undergoes a first-order transition, meaning that two distinct local minima at high magnetization and at zero magnetization become distinct global minima at the critical point and then swap their roles below. This leads to the phenomenon of phase coexistence. 

Our results show the following qualitative behavior of Code SW dynamics, which precisely mirrors the behavior of standard SW dynamics.
In contrast to local chains, Code SW dynamics is oblivious to the distinct global minima in the ordered phase of the 4D toric code. 
This can be understood through the fact that the uniform resampling of the cluster-update step automatically incorporates the symmetry of the model under a global logical operator, allowing the dynamics to traverse energy barriers. 
For such models, we can prove rapid mixing at any temperature.
In contrast, at the phase coexistence point of the 3-spin Curie-Weiss model, the two phases are not related by the intrinsic model symmetry and the SW dynamics gets trapped by the free-energy barrier. 
Consequently, we find torpid mixing at the first-order phase transition of this model.

\subsection{Relation to state of the art}
\label{sub:relation_to_state_of_the_art}

Sampling from commuting Hamiltonians, and in particular from the 2D toric code, has been investigated in recent, as well as older work. 
Already in 2009, \textcite{alicki_thermalization_2009} showed that a local Davies generator for the 2D toric code thermalizes quickly in a time upper bounded by $e^{O(\beta)}$, a scaling that matches the results for Glauber dynamics of the 1D Ising model \cite{levin_markov_2009}. 
Improving the scaling in $\beta$, recently, 
\textcite{ding_polynomial-time_2024} showed that a nonlocal Davies generator for the 2D toric code on $n$ qubits thermalizes in time at most $\min\{e^{O(\beta)}, \poly(n)\}$, leading to an efficient, nonlocal Gibbs sampler at arbitrary low temperatures. 
With an entirely different algorithmic approach based on a reduction to classical Gibbs sampling for 2-local Hamiltonians \textcite{hwang_gibbs_2025} showed that the Gibbs states of certain commuting Hamiltonians, including the 2D toric code can be prepared efficiently at arbitrary temperatures. 

Interestingly, two recent works identified distinct quantum algorithms preparing the Gibbs states of code Hamiltonians, which turn out to be special cases of $\Delta$-graphic codes.
First, \textcite{paez-velasco_efficient_2025} show that the Gibbs states of Hamiltonians which are \emph{poly-depth dual} to a collection of Ising chains can be prepared rapidly. 
A (stabilizer) Hamiltonian $H$ is poly-depth dual to an Ising Hamiltonian $H_I$ if there exists a polynomial-size circuit $U$ such that $H = U H_I U^\dagger$. 
Considering the symplectic matrix representations of $H$, $H_I$, and $U$, this definition is a special case of graphicness when restricted to low-depth bases, and hence all of their examples can be rapidly prepared using Code SW dynamics.
In addition to the 2D toric code, these include Haah's code \cite{haah_local_2011} and the $X$ cube \cite{weinstein_absence_2020} whose graphs are conjectured to be simple collections of cycles \cite{paez-velasco_efficient_2025}.

Second, \textcite{schmidhuber_hamiltonian_2025} showed that a Hamiltonian variant of Decoded Quantum Interferometry (DQI) can be used to efficiently prepare Gibbs states of stabilizer Hamiltonians satisfying a \emph{near-independence} property at any temperature. 
This property is defined as the number of bits encoded in the linear code generated by the rows of the symplectic binary matrix representation of the stabilizer Hamiltonian, capturing the number of linear dependencies between the checks. 
In particular, every nearly independent stabilizer code with $\Delta$ encoded bits, is $\Delta$-graphic, since after removing the linearly dependent checks, the resulting code is dual to a line graph. 
Thus, near-independence is a special case of graphicness. In contrast to CSW dynamics which only works for ferromagnetic models, however, HDQI can be used to prepare the Gibbs states of nearly independent signed stabilizer Hamiltonians as well. 

For the ferromagnetic case, $\Delta$-graphicness thus generalizes all known criteria for efficient Gibbs state preparation of code Hamiltonians, and moreover, CSW dynamics is significantly simpler than the known algorithms. 
None of these works have been able to generalize their results to models exhibiting thermally stable phases and, in particular, the 4D toric code.
$\Delta$-Cographicness seems to be essential in capturing such models, but it may well be the case that there are nontrivial $\Delta$-graphic codes with thermally stable phases as well.

For the 4D toric code, \textcite{bergamaschi_rapid_2025} showed that certain quasi-local block dynamics mix quickly \emph{within} a logical sector, but require exponential time to traverse its energy barrier. 
When applied to a specific input state---the maximally mixed state on the code space---this algorithm can be used to efficiently prepare the full Gibbs state of the 4D toric code. 
In contrast, the Code SW algorithm rapidly converges to the global Gibbs distribution from any initial state.

The Code SW algorithm is also closely related to the study of approximation algorithms for the Tutte polynomial of binary matroids~\cite{jackson_zero-free_2009,oxley_matroid_2011}. 
The mixing-time proof of \textcite{guo_random_2017} implies approximation algorithms for graphic as well as cographic matroids using duality properties of the Tutte polynomial. 
But while this algorithm just used the original SW algorithm for graphs, our algorithm shows how to explicitly sample from the random cluster and Gibbs distributions associated with a matroid. Our mixing-time bounds further imply extensions of those approximation results to matroids that are close to being graphic or cographic.
Conversely, \textcite{goldberg_polynomial-time_2013} show that the approximation problem for the Tutte polynomial can be solved for so-called regular matroids, leading us to conjecture that the Code SW algorithm also mixes quickly in this case.

Regarding our torpid mixing result for the $3$-spin Curie Weiss model, we highlight that we are able to achieve an $\exp(\Omega(n))$ mixing time lower bound for CSW dynamics. One can contrast this with SW dynamics where it was not until relatively recently that an $\exp(\Omega(n))$ lower bound was proven \cite{gheissari_exponentially_2018} using more sophisticated techniques than the original $\exp(\Omega(\sqrt{n}))$ lower bound by Gore and Jerrum \cite{gore_swendsenwang_1999} for the Curie-Weiss-Potts model. In the context of CSW dynamics for the 3-spin Curie-Weiss model, we find that the ordered phase has such high magnetization that an argument based solely on isolated vertices suffices to yield a $\exp(\Omega(n))$ mixing time lower bound.

\subsection{Next steps}
\label{sub:next steps}

Our work opens up many interesting questions about the mixing properties of CSW dynamics and generalizations to other models. 

First, we showed sufficient conditions for both rapid and torpid mixing, but precisely characterizing those regimes remains an open question. 
We conjecture that the only \emph{only obstacles} to rapid mixing are first-order phase transition points and that the Code SW mix rapidly everywhere else.
Given the fact that the Tutte polynomial can be efficiently computed for \emph{regular} matroids \cite{goldberg_polynomial-time_2013}, we also conjecture rapid mixing of CSW dynamics at any temperature in this case. 
On the other hand, at first order phase transition points, we believe all $p$-spin Curie-Weiss models for any $p$ exhibit torpid mixing.

In $p$-spin models at all other temperatures, there are no known Markov chains for sampling even from low-energy configurations, and hence, proving rapid mixing of CSW dynamics is an interesting open question. 
Similar results have been proven for the $q \ge 3$-state Curie-Weiss Potts model \cite{galanis_swendsen-wang_2019}. 
This would allow us to probe glassy behaviour which has been observed for the 3D plaquette model \cite{swift_glassy_2000}.

In which directions can CSW be generalized? Can we design parallel tempering or simulated annealing schedules for CSW dynamics which allow us to sample from the Gibbs state at first-order phase transitions \cite{bhatnagar_simulated_2016-1}? 
Can we extend CSW dynamics to frustrated systems~\cite{wang_antiferromagnetic_1989,kandel_cluster_1990}?
Interestingly, \textcite{borgs_efficient_2020} give an efficient sampling algorithm for the Potts model on the hypercubic lattice (that is different from SW) for any temperature, circumventing the bottlenecks at first-order transitions faced by Swendsen-Wang dynamics. 
Can similar results be achieved for code Hamiltonians?

In concurrent work, \textcite{pizzimenti_generalized_2025} propose an algorithm identical to CSW dynamics in the context of Potts lattice gauge theories, and provide numerics to support its rapid mixing at the phase transition point for the $\mathbb{Z}_2$ lattice gauge theory in 4D. This model is distinct from the 4D toric code, and does not appear to satisfy graphicness or cographicness. We leave as an open question the rapid mixing of CSW dynamics for the $\mathbb{Z}_2$ lattice gauge theory.
In fact, in this context, the random-cluster model has also received attention recently as a way to prove properties of the phase transitions in lattice gauge theories \cite{eldridge_cellular_2026}. 
We think it is an interesting open question to understand these results and potential generalizations in terms of the general code Hamiltonian framework we study here.

\subsection{Key ideas and technical overview}
\label{sub:key_ideas_and_technical_overview}

\paragraph{Rapid mixing}

Our proof builds on the framework of \textcite{guo_random_2017} and \textcite{ullrich_swendsen-wang_2014} for proving rapid mixing of the SW dynamics on the Ising model but requires a fundamentally new ingredient to handle codes beyond the graphic case.

Our proof for the $\Delta$-graphic case proceeds via a chain of three related Markov chains on subsets of checks indexed by $E \coloneqq [c]$:
\begin{enumerate}
    \item \emph{Code SW dynamics on the RC model}: Consider the Code SW chain where we focus our attention not on the spin configurations but on the \emph{check} configurations. 
    The stationary distribution of this chain is a \emph{random cluster} model on subsets of checks $S \subset E$  distributed as  $\phi(S) \propto p^{|S|}(1-p)^{|E \setminus S]}2^{k(S)}$ with $p = 1- e^{-2\beta}$. Here,  $k(S)$ is the dimension of the code defined by the checks in~$S$. 

    \item \emph{Single-check dynamics on the RC model}: The Markov chain which picks a random check and resamples it according to the Metropolis rule for $\phi$. 
    This can be seen as a ``slowed down'' version of CSW dynamics which updates a single, random check rather than all checks.

    \item \emph{``Code Jerrum-Sinclair (JS)'' chain on the even-cover model}: Consider the chain that attempts to sample from the distribution $\xi(B) \propto p^{|B|}(1-p)^{|E\setminus B|}$ supported on \emph{even covers} $B \subset E$, that is, subsets of checks where every variable is incident to an even number of checks. 
    This chain works by expanding the state space to include covers with the minimal number of defects needed to traverse between any two even covers.
\end{enumerate}
\textcite{ullrich_comparison_2013}'s proof straightforwardly extends to our setting, implying that the CSW dynamics for the RC model, and hence for the Gibbs state, mixes faster than single-check dynamics. 
If the model is graphic, we can reduce the even-cover model to an even-subgraph model. Using a coupling between the RC model and the even-cover model, we then construct a flow with low congestion to show that the single-check dynamics mixes rapidly, following Guo and Jerrum's \cite{guo_random_2017} argument.

It is in this last step that non-graphic codes pose a fundamental obstacle  to the efficiency of the code JS chain and thus the efficacy of the proof strategy of \textcite{guo_random_2017}.
This is because in order to even define the Code JS chain on even covers, their higher-dimensional connectivity requires that the state space be expanded by an exponential, rather than polynomial, factor, as is the case for graphic models.

Our key insight is that there is a \emph{dual} view, coupling the RC model to the \emph{syndrome distribution} of the code Hamiltonian. 
Analogously to the primal case, we can use this coupling to construct a good flow for the single-check dynamics, when the syndrome distribution possesses an approximate graphic form (our notion of approximate cographicness). 
This proves rapid mixing of the single-check chain, and thus the CSW chain as well.

\paragraph{Torpid mixing} In addition to our results on rapid mixing, we also show that CSW dynamics face an exponential bottleneck right at the \emph{first order} phase transition for the 3-spin Curie Weiss model \eqref{eq:curie weiss intro}.

As we vary the temperature of the Gibbs state, we see two types of extreme behaviors: At high temperature (low $\beta$), the Gibbs state will be highly concentrated at zero magnetization, and at low temperatures (high $\beta$), the Gibbs state will concentrate near the ground state, at high magnetization. These two phases, called the ``disordered'' and ``ordered'' phases, respectively, lead to drastically different behaviors in the \emph{random cluster} view, in terms of their density and structure.\footnote{The idea of viewing the slowdown of the SW chain in terms of the random cluster model originates in the work of \textcite{gore_swendsenwang_1999}, who use results on the emergence of the large component in a random graph to argue slow mixing. Instead of a random graph, the RC model for the CSW dynamics results in a random linear system.}

If $x$ is sampled from the ordered phase,  
the resulting system of equations in the cluster formation step of the CSW chain leads to random dense linear system. 
This is because almost every check is satisfied by $x$. 
When the number of equations $m$ in such a system exceeds the so-called ``clustering threshold'' at $m  \approx 0.818 n$, then the null space is known to cluster into regions that are well-separated, i.e., require flipping $\Omega(n)$ bits to reach one cluster from another, but individually well-connected, in the sense that any two elements in one cluster can be reached by flipping a small number of bits \cite{ach_molloy_clusters,mont_clustering}. 
Staying within the ordered phase (high magnetization states) is therefore equivalent to the existence of a single cluster, centered around zero.
Directly proving the existence of a unique cluster at this density appears to require significant effort, and we find it much easier to instead prove a bottleneck from the disorered to ordered phase.\footnote{Note that proving a bottleneck in one direction is equivalent to proving it for the other direction, since the chain satisfies detailed balance.}

Indeed, if $x$ is sampled from the disordered phase, the resulting system is both very structured and sparse.
It is structured because $E(x)$ contains only checks that cover three $+1$ spins and checks covering two $-1$ and one $+1$ spin. It is sparse because the total number of satisfied checks is much smaller than in the ordered phase. 
We argue that because of this this structure there are sufficiently many unchecked vertices in a RC sample from the disordered phase that only an exponentially small fraction of the solutions to the resulting linear system will be in the ordered phase. 
This shows the bottleneck from the disordered to the ordered phase of the model.

\section{Quantum Markov chain for code Hamiltonians}
\label{sec:quantum-classical reduction}

In this section, we describe a reduction from quantum to classical Gibbs sampling for Hamiltonians composed of commuting Pauli terms with positive coefficients (i.e., stabilizer Hamiltonians). Specifically, we show how Gibbs state preparation can be achieved by running a classical Markov chain for the Gibbs distribution of a corresponding linear code. Furthermore, we prove that the mixing time of the quantum chain is upper bounded by the mixing time of the classical chain.

\subsection{Preliminaries: Codes and their Gibbs states}

\subsubsection{Linear codes and stabilizer codes}

The \emph{linear code} defined by the parity check matrix $h \in \bin^{c \times n}$ is given by $C_h = \ker(h)$, the set of vectors $x$ such that $hx=0 \mod 2$. 

Let $P_1, \ldots, P_c \in \mc P_n$ be a set of commuting Pauli operators in the $n$-qubit Pauli group $\mc P_n$, and $\mc S = \langle P_1, \ldots, P_c \rangle \le \mc P_n$ be the stabilizer subgroup generated by Paulis $P_1, \ldots, P_c$. 
The \emph{stabilizer code} associated with $\mc S$ is given by 
\begin{align*}
  C_{\mc S} = \{\ket \psi \in (\mb C_2)^{\otimes n} : S \ket \psi = \ket \psi,  \forall S \in \mc S \}. 
\end{align*}

A convenient representation of Pauli operators is the \emph{stabilizer matrix representation}. 
Let $X,Y,Z$ be single-qubit Pauli operators and $X_i$ be the Pauli-$X$ operator acting on qubit $i$. Let $X(x) = \prod_{i \in [n]} X_i^{x_i}$ and $Z(z) = \prod_{i \in [n]} Z_i^{z_i}$ with $x,z \in \bin^n$.
We can thus represent every Pauli operator by a string $p = (x,z) \in \bin^{2n}$ via the mapping $P(p) \propto X(x)Z(z)$. For example, for a single qubit, $P(0,0) = \id$, $P(1,0) = X$, $P(0,1) = Z$ and $P(1,1) =Y$. The (multiplicative) Pauli group $\mc P_n$ without phases is therefore equivalent to the (additive) group $\mb Z_2^{2n}$ in the sense that for $x,y \in \bin^{2n}$, $P(x+y) \propto P(x)P(y)$.

The commutativity of Pauli operators is captured by the symplectic form  $\omega$ defined as follows
\begin{align*}
  [P(x),P(y)] =0 \, \Leftrightarrow \, \omega(x,y) \coloneqq x^T \omega y = 0 , \quad \omega = \begin{pmatrix}
    0 & \id_n\\ \id_n & 0 
  \end{pmatrix}.
\end{align*}

Given a stabilizer subgroup $\mc S$ generated by Pauli operators $P_1 = P(g_1), \ldots,  P_c = P(g_c)$ for $g_i \in \bin^{2n}$, we can therefore represent $\mc S$ by the \emph{stabilizer matrix} $g$ with rows $g_i$ as $\mc S = \mc P(g) \coloneqq \langle P(g_1), \ldots, P(g_c)\rangle $.
In fact, by linearity, only the row space of $g$ matters and we have $\mc S = \mc P(\row(g))$. 
Let $h = g\omega$. 
The commutant of $\mc S$ is given by $\mc C = \mc P(\ker(h))$ and contains both the stabilizer subgroup $\mc S$ itself and the logical operators $\mc L = \mc C /\mc S$ of the code $C_{\mc S}$.

Letting  $k=\dim(C_{\mc S})$, 
we can index a complete set of logical Pauli operators 
$\overline{P(l)}$ by $l \in \bin^{2k}$ 
 such that $\mc L = \langle \overline{P(l)} , l \in \bin^{2k} \rangle$. 
The \emph{syndrome} of $ C_{\mc S}$ is the column range $S_g= \col(g)$. It is the set of possible measurement outcomes when measuring the Pauli operators corresponding to the rows of $g$.
For every $s \in S$, let $h^{-1} s$ be the set of preimages of $s$ under $h$. We can associate a representative Pauli error $e_s = \min(h^{-1} s)$ given by the lexicographically first preimage and denote $E(s) = P(e_s)$.
It will be important that for any $e \in h^{-1} s$, $e_s + e \in \ker(h)$.

We write the projector onto the code space as $\Pi(g,0)  = 2^{-(n-k)}\sum_{S \in \mc S} S$, and the projector onto the syndrome-$s$ space as $\Pi(g,s) = E(s)\Pi(g,0)E(s)$. 

\subsubsection{Gibbs states of quantum and classical codes}
\label{ssub:gibbs_states_of_quantum_and_classical_codes}

Given a classical code with parity-check matrix $h \in \bin^{c \times n}$, we associate the energy function
\begin{align}
\label{eq:classical code hamiltonian}
  H(x) =  2|hx|,
\end{align}
where $|x | = \sum_i x_i$ is the Hamming weight of $x$. Note that this is equivalent to the form given in the introduction, \cref{eq:cch}, up to an additive shift. 
The corresponding Gibbs distribution is the probability distribution 
\begin{align}
\label{eq:gibbs distribution}
  \pi(x) \propto e^{- \beta H(x)}. 
\end{align}

Likewise, given a quantum stabilizer code with stabilizer matrix $g \in \bin^{c \times 2n}$, we associate the Hamiltonian 
\begin{align}
\label{eq:code hamiltonian}
  H = - \sum_{i \in [c]} P(g_i),
\end{align}
with Gibbs state 
\begin{align}
\label{eq:quantum gibbs state}
  \rho_\beta \propto e^{- \beta H}.  
\end{align}

\subsection{Lifting classical code sampling to quantum code sampling}

We now describe a quantum Markov chain for preparing Gibbs states of stabilizer codes (as in \cref{eq:quantum gibbs state}). The key idea is to lift a classical Markov chain $Q$, which samples errors from the Gibbs distribution of a corresponding classical code (of the form \cref{eq:gibbs distribution}), to a quantum chain whose fixed point is the Gibbs state.

For CSS codes, the formulation in the introduction---with two independent underlying chains---follows directly: 
the Pauli-$X$ and $Z$ terms decouple, so the Gibbs distribution factorizes into a product over $Z$ and $X$ errors. Hence, we can sample them independently. In contrast, for general stabilizer codes, the $X$ and $Z$ components are coupled, and we must consider the joint distribution over both types of errors.

To this end, let the stabilizer code (of dimension $k $) be specified by the stabilizer matrix $g \in \bin^{c \times 2n}$. 
Up to an additive shift, the  Hamiltonian \eqref{eq:code hamiltonian} can be rewritten as
\begin{align*}
  H = \sum_{s  \in S_g} 2|s| \Pi(g,s) ,
\end{align*}
which is diagonal in the syndrome-subspace projectors $\Pi(g,s)$. The Gibbs state is therefore given by
\begin{align*}
  e^{- \beta H} \propto  \sum_{s \in S_g} e^{-2\beta |s| } \Pi(g,s) .
\end{align*}

Let $\pi$ be the Gibbs distribution associated with $h \coloneqq g \omega$ at inverse temperature $\beta$. 
We will lift a classical Markov chain that is irreducible, aperiodic and reversible with respect to $\pi$  with transition matrix $Q(\cdot, \cdot)$ to a quantum Markov chain converging to the Gibbs state $\rho_\beta$. 
We restrict ourselves to Markov chains such that $Q(x, \cdot)$ is invariant under a shift by a logical operator at \emph{every} step. Formally,
\begin{equation} \label{eq:implicit_logical}
    Q(x, y) = Q(x, y + \ell),
\end{equation}
for any $\ell \in \{0,1\}^{2n}$ representing a logical operator, i.e., $P(\ell) \in \mc L$. Indeed, we show that CSW dynamics satisfies this condition in \cref{lemma:csw_qcreduction}.

\begin{algorithm}[h]
\caption{Quantum code Gibbs sampling
\label{alg:quantum sampler}}
  \begin{algorithmic}[1]
    \Require Stabilizer matrix $g$, classical Markov chain $Q(\cdot, \cdot)$, initial state $\ket{\phi_0}$, initial noisy codeword $x_0$.
    \State Let  $\ket \phi \leftarrow \ket {\phi_0}, x \leftarrow x_0$, cutoff time $T$.
    \State Iterate $T$ times: \\ 
    \begin{enumerate}[label=\roman*.]
       \item Measure the stabilizers $\mc S$ on $\ket \phi$, yielding syndrome $s \in S_g$, and post-measurement state $\ket {\phi_s}$.
      \item Sample $y \leftarrow Q(x, \cdot) $, and apply $P(y)E(s)$, letting $\ket \phi \leftarrow P(y)E(s)\ket {\phi_s}$, $x \leftarrow y$. 
     \end{enumerate} 
    \Ensure $\ket \phi $
  \end{algorithmic}
\end{algorithm}

\begin{theorem}[Convergence to the Gibbs state]
\label{thm:convergence to gibbs state}
  The outputs $\ket \phi$ sampled from \cref{alg:quantum sampler} satisfy 
  \begin{align*}
    \lim_{T \rightarrow \infty} \Eb_{ \phi} \proj \phi \propto e^{- \beta H} ,
  \end{align*}
  for any initial state $\ket {\phi_0}$.
\end{theorem}
\begin{proof}
We begin by  observing that at any step of the chain, the post-measurement  state is given by an erroneous logical state $E(s)\overline{\ket \varphi}$ for some $s \in S_g \subset \bin^c$, where  $\overline{\ket \varphi} = \Pi(g,0)\overline{\ket \varphi}$ is a code state. 
Now, observe that a sample  $y \leftarrow \pi$ can be decomposed as 
\begin{align}
  \label{eq:decomposition y}
  y = d  + e + f 
\end{align}
in terms of a logical Pauli, a representative error, and a stabilizer represented by $d, e, f \in \bin^{2n}$, respectively.
We can thus write 
\begin{align*}
   P(y) \propto E(s) \overline{P(\ell)} S,
\end{align*}  
up to a global phase for a syndrome $s = he$, a logical $\ell \in \bin^{2k}$ such that $P(d) = \overline{P(l)}$, and a stabilizer $S = P(f) \in \mc S$. 

Applying $E(s) $ moves the state to the logical subspace, i.e., the $0$-syndrome subspace. 
Due to \cref{eq:implicit_logical}, $\ell$ is uniformly random, and applying $P(\ell)$ then twirls any logical state $\overline{\ket \varphi}$ as 
\begin{align*}
  \Eb_{\ell \in \bin^{2k}} \overline{P(\ell)} \, \overline {\proj{\varphi}}  \,\overline{P(\ell)} = \frac 1 {\tr[\Pi(g,0)]}\Pi(g,0) = \frac 1 {2^k}\Pi(g,0). 
\end{align*}
Therefore, the expected output of the algorithm only depends on the syndrome distribution. Since $Q^T(x_0, \cdot) \rightarrow \pi$, we compute the limit for an independent sample $ y \leftarrow \pi$ and any fixed logical state as 
\begin{align*}
  \rho & \propto
   \sum_y e^{-2 \beta |hy|} P(y) E(s) E(s) \overline{\proj{\varphi}} E(s) E(s) P(y)\\
  & =   \sum_{t \in S_g} e^{-2\beta |t|} E(t) \left[  \Eb_{\ell \in \bin^{2k}} \overline{P(\ell)}  \,\overline {\proj{\varphi}} \, \overline{P(\ell)}\right] E(t)\\
  & = \frac1 {2^{k}} \sum_{t \in S_g} e^{-2\beta |t|} E(t)\Pi(g,0) E(t) \\
  & = \frac1 {2^{k}} \sum_{t \in S_g} e^{-2\beta |t|} \Pi(g,t) \propto e^{- \beta H}
\end{align*}

\end{proof}

We now relate the mixing times of the quantum and classical chains. 
To state the mixing time bound, let $P$ be the transition matrix for an an ergodic (classical) Markov chain on~$X$ with stationary distribution~$\pi$. We define the mixing time 
\begin{align*}
  \tau(P)\coloneqq \min_t \left[\max_{x \in X} \sum_{y} |P^t(x,y)  - \pi(y)| \le e^{-1} \right]. 
\end{align*}
We furthermore define the mixing time of \cref{alg:quantum sampler} which outputs a state~$\ket \phi$ after time~$T$ as 
\begin{align*}
  \tau_q \coloneqq \min_T \left[ \max_{\ket{\phi_0} \in (\mb C^2)^n} \norm{\Eb_\phi \proj{\phi} - \rho_\beta}_{\tr}  \le e^{-1} \right] . 
\end{align*}

\begin{lemma}[Coupling of quantum and classical chains] \label{lemma:qccoupling}
  The mixing times of $Q$ and \cref{alg:quantum sampler} satisfy $\tau_q \le \tau(Q)$ if $Q$ satisfies  \eqref{eq:implicit_logical}. 
\end{lemma}

\begin{proof}
  Due to \eqref{eq:implicit_logical}, a single sample from the chain $y \leftarrow Q(x_0, \cdot)$ is equivalent to a sample $y + \ell$ for $y \leftarrow Q(x_0, \cdot)$ and $\ell$ uniformly random s.t.\ $P(\ell) \in \mc L$. Therefore, at time $t$ the average state of the algorithm is given 
  by 
  \begin{align*}
    \rho_t = \frac{1}{2^k} \Eb_{y \leftarrow Q^t(x_0,\cdot)} P(y) \Pi(g,0) P(y),
  \end{align*}
where we use the fact that twirling any state with a random logical operator results in a maximally mixed state over the codespace.
We can further write the equilibrium state $\rho_\beta \propto e^{-\beta H}$ as  
\begin{align*}
  \rho_\beta = \frac{1}{2^k}\Eb_{y \leftarrow \pi} P(y) \Pi(g,0) P(y),
\end{align*}
by applying the proof of \cref{thm:convergence to gibbs state} in reverse. 
Therefore, 
\begin{align*}
 \norm{\rho_t- \rho_\beta}_{\tr} & =  \norm{\sum_{y} \left(Q^t(x_0,y) - \frac 1 Z e^{- 2\beta |hy|}\right)   P(y) \frac{\Pi(g,0)}{2^k} P(y)}_{\tr}\\ 
 & \le \sum_y |Q^t(x_0,y) - \frac 1 Z e^{- 2\beta |hy|}| \le 1/e,
\end{align*}
where $Z = \sum_y e^{- 2\beta|hy|}$ and we used that the trace norm of a quantum state is equal to 1. 
\end{proof}

\subsection{Quantum Markov chains beyond code Hamiltonians}

We note that \cref{alg:quantum sampler} is not restricted to stabilizer Hamiltonians, but will also be correct for \emph{signed} stabilizer Hamiltonians. 
Such Hamiltonians are parameterized by a stabilizer matrix $g \in \bin^{c \times 2n}$ as well as a vector $t \in \bin^c$ as 
\begin{align*}
  H= - \sum_{i \in [c]} (-1)^{t_i} P(g_i). 
\end{align*}
For this signed case---analogous to antiferromagnetic Ising Hamiltonians---it is well known, however, that sampling is intractable. 
In fact, an efficient sampler would be able to solve \maxcut, an \np-complete problem. 
Therefore, we also do not expect an efficient quantum sampler for this case. 

We can also conceive of analogous dynamics for  Hamiltonians which are sums of arbitrary---potentially non-commuting---Pauli operators. 
Importantly, the distribution $\pi$ is still well-defined in this case and the energy function just corresponds to the number of Hamiltonian terms an ``error'' anticommutes with. 
However, in this case, syndrome extraction cannot be done for all terms simultaneously and the measurement will always project into the joint eigenspace of a subset of the Hamiltonian terms, giving a non-convergent chain. 
We believe it is an interesting future direction to explore the dynamics that arise from \cref{alg:quantum sampler} for different measurement protocols of the Hamiltonian terms, however. 
Some examples that could be interesting are: 
(i) sequentially measure random maximal subsets of commuting terms and 
(ii) weakly measure the Hamiltonian terms.


\section{Code Swendsen-Wang dynamics and mixing time}
\label{sec:classical}

In this section, we formally define the (classical) Code SW chain and show that it converges to the correct Gibbs distribution (\cref{sub:sw codes}). Before we do this, we recap the SW dynamics for the Ising model (\cref{sub:recap ising}).
We then introduce structural conditions on parity check matrices---$\Delta$-graphic and $\Delta$-cographic---under which we prove that the Code SW chain mixes rapidly (\cref{sub:mixing sw}). 
Finally, we prove that the Code SW chain can exhibit torpid mixing at first order phase transitions (\cref{sub:slow mixing}).

\subsection{Recap: Swendsen-Wang dynamics for the Ising model}
\label{sub:recap ising}

Let $G = (V,E)$ be a graph. We consider spins on the vertices with configuration space $\Sigma = \pmbin^V$. 
The (ferromagnetic) Ising model on $\Sigma$ is described by the energy function 
\begin{align*}
  H(\sigma) = - \sum_{(u,v) \in E} \sigma_u \sigma_v. 
\end{align*}

Recall the Swendsen-Wang (SW) chain for sampling the Gibbs distribution $\pi(\sigma) \propto e^{-\beta H(\sigma)}$: For a given configuration of spins $\sigma$, define the set of satisfied edge constraints
\begin{align*}
  E(\sigma) &\coloneqq \{ \{u,v\} \in E: \, \sigma(u) = \sigma(v) \}. 
\end{align*} Then the SW chain iterates the following two steps, starting from a spin configuration $\sigma \in \Sigma$. 
\begin{enumerate}
  \item \emph{Cluster formation:} For each $e \in E(\sigma)$ include it in $A $  with probability $p = 1- e^{-2\beta}$, giving $A \subset E$.

  \item \emph{Cluster update:} For every connected component $C \subset V $ of $A$ let  $\tau(C) = +1$ with probability $1/2$    and $-1$ otherwise, giving a new state $\tau \in \Sigma$.
\end{enumerate}

This chain is motivated by a correspondence, discovered by Fortuin and Kasteleyn \cite{fortuin_random-cluster_1972}, between the Gibbs measure $\pi(\sigma)$ and the random cluster (RC) measure 
\begin{align*}
  \phi(A) \propto \left( \frac p {1-p} \right)^{|A|} 2^{k(A)} ,
\end{align*}
on subgraphs $A \subset E$, 
where we denote the number of connected components of $A$ by $k(A)$. 
In particular, the correspondence is given by the joint distribution
\begin{align*} 
  \mu(\sigma, A)  &\propto \left(\frac p {1-p}\right)^{|A|}\id(A \subset E(\sigma)), 
\end{align*}
where we write $\id(e)$ for the indicator of an event $e$. 

\begin{claim}[Coupling of Ising and RC model] \label{claim:margin_gibbs}
$\mu$ defines a proper coupling between $\pi$ and $\phi$, with $p=1-e^{-2\beta}$, i.e., 
\begin{align*}
\sum_{A \subset E} \mu(\sigma, A) = \pi(\sigma),  \qquad 
   \sum_{\sigma \in \Sigma} \mu(\sigma, A) = \phi(A).
\end{align*}
\end{claim}
\begin{proof}
To show the claim, we observe that 
\begin{align*}
  \mu(\sigma, A) & \propto \prod_{(i,j)=e \in A} \left[ p \id(\sigma(i)=\sigma(j)) \right] \prod_{e \notin A} (1-p)  \\
  & \propto \prod_{e=(i,j) \in E} \left[p \, \id(\sigma(i) = \sigma(j)) \id( e \in A) +(1-p) \id(e \notin A)  \right].  
\end{align*}

Setting $p=1-e^{-2\beta}$, we then have
\begin{align*}
    \sum_{A \subset E} \mu(\sigma, A) & \propto \prod_{(i,j) \in E} \left( (1-e^{-2 \beta}) \id(\sigma(i) =  \sigma(j))  +e^{-2 \beta}  \right) \\
    &= \prod_{(i,j) \in E \; : \; \sigma(i) \neq \sigma(j)} e^{-2\beta} \\
    & \propto e^{-\beta H(\sigma)},
\end{align*}and

  \begin{align*}
      \sum_{\sigma \in \Sigma} \mu(\sigma, A) &\propto p^{|A|}(1-p)^{|E\setminus A|} \sum_{\sigma} \prod_{(i,j) \in A} \id(\sigma(i) = \sigma(j)) \\
      &= p^{|A|}(1-p)^{|E\setminus A|} 2^{k(A)} \\
      & \propto \left( \frac p{1-p}\right)^{|A|} 2^{k(A)}, 
  \end{align*}
  which completes the proof.
\end{proof}

Each step of the SW chain simply draws conditional samples from this joint measure $\mu(\sigma, A)$. In particular, the cluster formation step (step 1) is equivalent to drawing a sample $A \leftarrow \mu(\sigma, \cdot)$, and the cluster update step (step 2) is equivalent to drawing a sample $\tau \leftarrow \mu(\cdot, A)$. 

From this, we can immediately see that the SW dynamics has the distribution $\pi$ as its unique fixed point, since its transition matrix $P(\sigma, \tau) = \sum_A P(\sigma, A)P(A,\tau)$ satisfies detailed balance:
\begin{align}
\label{eq:detailed balance sw}
  \frac{P(\sigma, \tau)}{P(\tau,\sigma)} = \frac{ \sum_A  \mu(A|\sigma)\mu(\tau|A)}{\sum_A  \mu(A|\tau)\mu(\sigma|A)} = \frac{ \pi(\tau)}{\pi(\sigma)},
\end{align}
where we have simply used the law of conditional probabilities $\mu(A|\sigma) = \mu(\sigma,A)/\pi(\sigma)$ and likewise for $\mu(\sigma|A)$.
To complete the argument, observe that the chain is aperiodic and ergodic since after step 1 not a single cluster survives with finite probability, and therefore, in step 2, an arbitrary state can be reached.

\subsection{Code Swendsen-Wang dynamics}
\label{sub:sw codes}

To generalize the SW dynamics to classical codes, we observe that the Ising model on a graph $G = (V,E)$ defines a particular linear code. 
To see this, notice that every edge of the graph defines a pairwise parity check constraint. If we interpret the edge-vertex incidence matrix $h \in \bin^{|E| \times |V|}$ of $G$ as a parity check matrix, then the corresponding code $C_h$ of this parity check matrix is the repetition code, $\{0^n, 1^n \}$, if the graph is fully connected. 
Viewing the Ising model this way, there is nothing special about the relationship between the SW chain and the Ising model. In fact, we are free to choose \emph{any} parity check matrix $h$ of a linear code and define an analogous SW dynamics for that code. 

More precisely, in the following we consider a code with parity check matrix $h \in \bin^{c \times n}$,
and the  configuration spaces of variables $X = \bin^n$ and (indices of) checks $E = [c]$. 
Our goal is to sample from the Gibbs distribution  $\pi(x) \propto e^{-2 \beta |hx|}$, see \cref{eq:gibbs distribution}.  Toward this end, let $\supp(x) = \{i \in [n]: x_i = 1 \}$ be the support of $x$ and define the set of satisfied checks
\begin{align*}
  E(x) &\coloneqq E \setminus \supp( h x) 
\end{align*}
and the parity-check matrix $h_A$ as the row-submatrix of $h$ indexed by $A \subset E$.
The Code Swendsen-Wang (SW) algorithm is given in \cref{alg:code sw}. 
\begin{algorithm}[h!]
\caption{The Code SW algorithm}
\label{alg:code sw}
  \begin{algorithmic}[1]
    \Require Parity check $h$, inverse temperature $\beta$, initial state $x$, cutoff time $T$.
    \State Let $x \leftarrow x_0, A \leftarrow \emptyset$.
    \State{Iterate $T$ times: 
    \begin{enumerate}[label=\roman*.]
    \item
     \emph{Cluster formation:} For every $e \in E(x)$, include it in $A$ with probability $p = 1- e^{- 2\beta}$, giving $A \subset E$.
    \item \emph{Cluster update:} Pick $y \in \ker (h_A)$ uniformly at random and let $x \leftarrow y, A \leftarrow \emptyset$.
    \end{enumerate}
    }
    \Ensure $x$
  \end{algorithmic}
\end{algorithm}

To see that the Code SW algorithm converges to its unique fixed point $\pi$, we observe that it, too, works by coupling to the random cluster model on codes, defined for $A \subset E$ as
\begin{align*}
   \phi(A) \propto \left(\frac p{1-p}\right)^{|A|} 2^{k(A)}, 
\end{align*}
where $k(A)$ is the dimension of $\ker(h_A)$. This is achieved via a generalized FK measure 
\begin{align} 
  \mu(x, A)  &\propto \left( \frac{p}{1-p} \right)^{|A|} \id(A \subset E(x))
\end{align}
at $p = 1- e^{-2\beta}$.

\begin{lemma}[Correctness of Code SW dynamics]
  The Code SW algorithm (\cref{alg:code sw}) converges to $\pi$ as $T \rightarrow \infty$ for any initial state $x_0$. 
\end{lemma}
\begin{proof}
  Clearly, the chain is aperiodic and ergodic, since after a single iteration of the algorithm any state can be reached since $\Pr(A = \emptyset)>0$  for $\beta<\infty$. 
  To show that it satisfies detailed balance, we use \cref{eq:detailed balance sw} and all that remains is to show that the transition probabilities of step 1 and 2 in \cref{alg:code sw} are 
\begin{align*}
  P(x,A) = \mu(A|x),  \qquad
   P(A,y) = \mu(y|A).
\end{align*} 
To see this, observe that $\mu$ is a valid coupling, i.e., 
\begin{align*}
  \sum_{ x \in X} \mu(x,A) = \phi(A) \; \text{ and }  \sum_{A\subset E} \mu(x,A)
  & = \pi(x).
\end{align*}
This can be shown via a proof identical to \cref{claim:margin_gibbs}, observing that 
\begin{align*}
  \mu(x,A) \propto \prod_{e \in E} \left(  p \, \id( h_e x= 0) \id(  e \in A) + (1-p) \id(e \notin A)\right).
\end{align*}

It remains to consider the case of zero temperature.
In this case, we can see that the algorithm converges rapidly: 
In the  cluster-formation step all previously satisfied checks are kept. 
In the cluster-update step, if there is an unsatisfied check prior to the update, this check will be satisfied with probability $1/2$ after the update. 
Therefore the number of unsatisfied checks converges to zero exponentially fast. 
Once no unsatisfied checks remain, the cluster update step is a sampler from the zero-temperature Gibbs state. 
\end{proof}

The Code SW chain is also compatible with the quantum-to-classical reduction given in \cref{alg:quantum sampler}. 

\begin{lemma} \label{lemma:csw_qcreduction}
    Let $\tau_q$ be the mixing time of the quantum Markov chain in \cref{alg:quantum sampler} instantiated with the CSW chain and $\tau_{CSW}$ the mixing time of the CSW chain. Then $\tau_q \leq \tau_{CSW}$. 
\end{lemma}
\begin{proof}
    To apply \cref{lemma:qccoupling}, it suffices to show that the CSW chain satisfies the logical invariance property in \eqref{eq:implicit_logical}. Indeed, in the cluster formation step, the set $A$ is always a subset of $E$, the parity checks of the code. Thus, $\ker (h_A)$ always contains the subspace of logical operators, i.e., vectors $\ell \in \{0,1\}^{2n}$ s.t.\ $P(\ell) \in \mc L$. Therefore, a random sample from the subspace $\ker (h_A)$ is invariant under a shift by $\ell$.
\end{proof}

\subsection{Rapid mixing of the Code Swendsen-Wang dynamics}
\label{sub:mixing sw}

In this subsection, we prove that the Code SW dynamics is rapidly mixing for codes that are $\Delta$-graphic or $\Delta$-cographic (see below for definition). Our proof proceeds in two steps: first, we adapt Ullrich's proof that SW dynamics mixes quicker than single-bond dynamics to the setting of linear codes; second, we show that single-check dynamics -- the analogue of single-bond dynamics -- is rapidly mixing for $\Delta$-graphic and $\Delta$-cographic codes.

\subsubsection{Comparison to single-check updates}

For the proof, we will first switch to the RC view of the SW algorithm. 
To this end, observe that if we halt the algorithm after the cluster formation step (step 1), the sampled subset of checks will be distributed according to the RC model $\phi$  (as $T \rightarrow \infty$).

Therefore, to prove mixing times of the SW algorithm for the Gibbs distribution, it is sufficient to prove a mixing time bound for the convergence to the RC model instead. 

We start with a comparison to the standard (lazy) single-check-update Metropolis dynamics in which with probability $1/2$ no change is made and otherwise, an update $B = A \oplus e$ is proposed for uniformly random $e \in E$. This proposal is accepted with probability
\begin{align*}
\Pr(\text{Accept}) = \min\left\{ 1, \frac {\phi(B)}{\phi(A)} \right\} ,
\end{align*}
and rejected  otherwise.
Let 
\begin{align*}
  P_{\text{Metropolis}}(A,B) = \begin{cases}
    \frac 1 {2|E|} \min\{ 1, \frac {\phi(B)}{\phi(A)} \} & |A \oplus B| = 1\\ 
    1 - \frac 1 {2|E|} \sum_{e \in E} \min\{ 1, \frac{\phi(A \oplus e )}{\phi(A)} \} & A = B \\
    0 & \text{ else}
  \end{cases} 
\end{align*}
be the corresponding transition matrix, and likewise $P_{\text{SW}}$ be the transition matrix of the SW process for the RC model.

\begin{lemma}[SW is faster than Metropolis dynamics]
\label{lem:sw vs sc}
The SW dynamics of the RC model is faster than Metropolis dynamics (up to a constant), i.e.,
\begin{align*}
    \tau(P_{\text{Metropolis}}) \ge \frac 12 \tau(P_{\text{SW}}).
\end{align*}
\end{lemma}
The proof of \cref{lem:sw vs sc} is an adaptation of the result of \textcite{ullrich_swendsen-wang_2014} to our setting, which we give in Appendix~\ref{app:sw vs sc}.

\subsubsection{Rapid mixing of single-check dynamics}

Here we show that single-check dynamics for the RC model of a parity check matrix $h$ mixes rapidly if $h$ has an approximate ``graphic'' or ``co-graphic'' representation.

\begin{definition} 
\label{def:graphic cographic}
  Let $h \in \{0,1\}^{c \times n}$ be a parity check matrix. We say that $h$ is 
  \begin{itemize}
    \item $\Delta$-graphic if there exists an edge-vertex incidence matrix $g \in \{0,1\}^{c \times m}$ of a graph on $m \leq O(n)$ vertices such that $\ker(g^T) \supset \ker(h^T)$ and $\dim(\ker(g^T))-  \dim (\ker(h^T)) \le \Delta$.\footnote{The restriction to $m=O(n)$ is just for convenience in stating our formal results. In fact, any $m=\poly(n)$ will suffice for rapid mixing.}
    \item $\Delta$-cographic if a column-generator matrix $h^\perp$ of $\col(h)^\perp$ is $\Delta$-graphic. 
  \end{itemize}
  In the first case, we call the edge-vertex incidence matrix $g$ the \emph{primal coupling} to $h$. In the second case, we call the edge-vertex incidence matrix $g$ that is the primal coupling to $h^\perp$ the \emph{dual coupling} to $h$.
\end{definition}

Which codes admit primal or dual couplings?

\begin{claim}
    The parity check matrix $h$ of any linear code with independent checks is $0$-graphic.
\end{claim}
\begin{proof}
    Let $g$ be the edge-vertex incidence matrix of a line graph. Then $\ker(g^T)$ and $\ker(h^T)$ are both 0-dimensional. 
\end{proof}

\begin{claim}
    The X and Z parity check matrices $h$ of the 2D toric code are $0$-graphic
\end{claim}
\begin{proof}
    Consider the edge-vertex incidence matrix $g \in \{0,1\}^{c \times c}$ for a cycle graph. Then $\ker(g^T)$ is spanned by the all $1$'s vector, which matches exactly with the even covers of the 2D toric code. {}
\end{proof}

\begin{claim}
    The X and Z parity check matrices $h$ of the 4D toric code are $4$-cographic.
\end{claim}
\begin{proof}
    Let $g$ be the edge-vertex incidence matrix of the 4D torus. Then $g^T h = 0$ because $g^T$ and $h$ are boundary maps of a chain complex. Therefore, $\ker(g^T) \supset \col(h) = \ker((h^\perp)^T)$. On the other hand, a set of generators for the even subgraphs of the 4D lattice are the trivial loops (boundaries of faces) and nontrivial loops (crossing the boundary of one of the dimensions). There are four generators for the latter, corresponding to each dimension, so $\dim(\ker(g^T)) - \dim(\col(h)) = 4$. 
\end{proof}

To show rapid mixing of the single-check dynamics of the RC model for these models, we will use the method of \emph{flow congestion}. 
To introduce it, we first introduce the concept of a path in the state space. 
Let the state space of our Markov chain with transitions $P$ be $\Omega$, and the \emph{graph} of $P$ be defined by the edges $\mc E(P) \coloneqq \{(U,V) \in \Omega \times \Omega: P(U,V)> 0\} $.  
A path 
\begin{align*}
  \gamma = (\gamma_0, \gamma_1,\ldots, \gamma_{L(\gamma)}) \subset \Omega^{*},
\end{align*}
of length $L(\gamma) \in \mb N$ is a sequence of states $\gamma_i \in \Omega$ which are connected by transitions of the Markov chain, i.e., $(\gamma_i, \gamma_{i+1}) \in \mc E(P)$. 
Let $\Gamma(I,F) \coloneqq \{\gamma \in \Omega^*: \gamma_0 = I, \gamma_{-1} = F \}$ be the set of paths from $I$ to $F$ and $\Gamma = \bigcup_{I,F} \Gamma(I,F)$. 
\begin{definition}[Flows and canonical paths]
A \emph{flow} with respect to a probability distribution $p$ on $\Omega$ is a function $f: \Omega^{*} \rightarrow [0,1]$ assigning a weight $f(\gamma)$ to every path such that 
\begin{align*}
  \sum_{\gamma \in \Gamma(I,F)} f(\gamma) = p(I)p(F)
\end{align*}
for all $I \neq F \subset E$.
A set of \emph{canonical paths} for $p$is a flow for which there is only a single ``canonical'' path between every pair $(I,F)$, i.e., $|\Gamma(I,F)| =1$.
In this case $f(\gamma) = p(I)p(F)$ for $\gamma \in \Gamma(I,F)$. 
\end{definition}

Rapid mixing is equivalent to the existence of flows with low \emph{congestion}, where the congestion of a set of flows $f$ is given by 
\begin{align*}
  \rho(f) \coloneqq \max_{(Z,Z') \in \mc E(P)} \frac 1 {p(Z) P(Z, Z')} \sum_{\gamma \in \Gamma, (Z,Z') \in \gamma}  f(\gamma) L(\gamma),
\end{align*}

A fundamental result in the analysis of mixing times of Markov chains is that there is a flow such that the mixing time is captured by the congestion of that flow. 
\begin{theorem}[\cite{sinclair_improved_1992,guruswami_rapidly_2016}] \label{thm:mix time}
For a lazy, ergodic, reversible Markov chain $P$, we have 
\begin{align*}
\Omega\left(\inf_f \rho(f)\right) \le  \tau(P) \le  \ln\left(\frac{2e}{\min_{S \in \Omega} p(S)} \right) \rho(f).
\end{align*}
\end{theorem}

The mixing time of a Markov chains is therefore equivalent to the congestion of some flow. 
To show rapid mixing, we will construct a ``good'' flow for the RC model $\phi$.

\begin{theorem}[Existence of a good flow for the RC model] \label{thm:main theorem}
  Suppose that $h \in \bin^{c \times n}$ is $\Delta$-graphic or $\Delta$-cographic. Then there is a flow $F$ for the RC model $\phi$ with congestion
  \begin{align*}
    \rho(F) \le O(c^2 2^{2\Delta+5} n^4). 
  \end{align*}
\end{theorem}

Combining this with \cref{thm:mix time}, we arrive at a polynomial mixing-time bound. 
\begin{corollary}[Rapid mixing of Code SW dynamics]
  Suppose that $h \in \bin^{c \times n}$ is $\Delta$-graphic or $\Delta$-cographic, for $\Delta \leq O(\log n)$. Then the Code SW chain for $h$ mixes in $\poly(n, \beta)$.
\end{corollary}
\begin{proof}
By \cref{thm:mix time} and \cref{thm:main theorem},
\[
\tau(P_{\mathrm{SW}})
\le
c^2 2^{O(\Delta)}n^4
\log\!\left(\frac{2e}{\phi_{\min}}\right).
\]
Since
\[
\phi(A)=Z^{-1}p^{|A|}(1-p)^{c-|A|}2^{k(A)},\qquad Z\le 2^n,
\]
we have
\[
\phi_{\min}\ge 2^{-n}\min\{p,1-p\}^c.
\]
Thus, for \(p=1-e^{-2\beta}\),
\[
\tau(P_{\mathrm{SW}})
\le
c^2 2^{O(\Delta)}n^4
\left(
2+n\log 2+
c\max\left\{
2\beta,\log\frac{1}{1-e^{-2\beta}}
\right\}
\right),
\]
which is polynomial in \(n\) and $\beta$ for \(\Delta=O(\log n)\).
\end{proof}

We prove \cref{thm:main theorem} by coupling the RC model on $h$ to a generalization of the even subgraph model, which we dub the \emph{even cover model}.
The even cover model is defined by the distribution 
\begin{align*}
  \xi_{h,p}(A) \propto \left(\frac{p}{1-p}\right)^{|A|} \id( 1_A \in \ker(h^T)),
\end{align*}
where $1_A $ is the indicator of $A \subset E $ with $\supp(1_A) = A$. 
In words, this is a weighted distribution over subsets of checks such that every vertex is incident to an even number of checks. 
The following lemma makes transparent the connection between the even cover model and the RC model. We will make the dependence of $\phi$ on $p$ explicit by writing $\phi_p = \phi$ whenever it is needed.

\begin{lemma}[Coupling of even covers and RC]
\label{lem:even covers rc coupling}
  There are two couplings of the even-cover model to the random cluster model. 
  Let $h^\perp$ be a column-generator matrix of $\col(h)^\perp$.
  \begin{itemize}
    \item (primal lift) Let $A \sim \xi_{h,p/2}$, and add each $e \in E \setminus A$ iid.\ with probability $(p/2)/(1-p/2)$ to obtain $B$. Then $B \sim \phi_p$. 
    \item (dual lift) Let $A \sim \xi_{h^\perp,(1-p)/(2-p)}$, and add each $e \in E \setminus A$ iid.\ with  probability $1-p$ to obtain $B$. Then $E\setminus B \sim \phi_p$. 

  \end{itemize}
  We call $\xi_{h,p/2}$ the \emph{primal coupling} to $\phi$ and $\xi_{h^\perp,(1-p)/(2-p)}$ the \emph{dual coupling} to $\phi$.
\end{lemma}
We prove \cref{lem:even covers rc coupling} in \cref{sec:coupling_to_syndromes_and_even_covers}.

Our high-level strategy for constructing flows for the RC model is similar to the strategy of \textcite{guo_random_2017}. 
They begin with a parity check matrix $g$ that is an edge-vertex incidence matrix and consider the primal coupling $\xi_{g,p/2}$. 
In this case, the primal coupling is a weighted distribution over \emph{even subgraphs} of the original graph, i.e., subgraphs that have even degree on every vertex.
Then, they use an idea due to \textcite{jerrum_polynomial-time_1993}: 
they enlarge the state space to also include subgraphs with exactly \emph{two} defects---odd degree vertices---called \emph{worms}.
This enlargement allows any two even subgraphs $S,T \subset E$ to be connected by sequences of single edge-flips that sequentially unwind the loops in $S \oplus T$. 
Any intermediate state of these paths is a worm state, i.e., has at most two defects. 
These paths can be ``lifted'' to a good flow for the RC model by \emph{mimicking} edge additions and deletions from unwinding even subgraphs to additions and deletions in the RC model. 

Our proof strategy for constructing good flows for $\Delta$-graphic and $\Delta$-cographic $h$ proceeds by adapting the above strategy in two ways. 
First, instead of just starting from the primal coupling to even covers $\xi_{h,p/2}$ of $h$, we can also start from the dual coupling $\xi_{h^\perp,(1-p)/(2-p)}$ to even covers of $h^\perp$.  
Second, when $\Delta$ is nonzero, the even covers of $h$ or $h^\perp$ are \emph{strictly contained} in the even subgraphs $\Omega_0 \coloneqq \ker(g^T)$ of a graph defined by the respective parity check $g$ from \cref{def:graphic cographic}, i.e.,  
$\ker(h^T) \subsetneq \Omega_0$ if using the primal coupling or $\col(h) \subsetneq \Omega_0$ if using the dual coupling. 
Because we only know how to efficiently sample from \emph{all} even subgraphs, but not a subset thereof, this  means that we now need to instead lift paths for this extended space. 
We show that this difference only leads to an increase in the mixing time of $2^{2\Delta}$.

For the formal argument, let 
\[
p_{\uparrow} = p/2, \quad p_{\downarrow} =(1-p)/(2-p)
\] 
be the weight parameters for our primal ($\uparrow$) and  dual ($\downarrow$) coupling to $\phi$. 
In what follows we will write $p_{\uparrow / \downarrow}$ to mean $p_{\uparrow}$ if doing a primal lift and $p_{\downarrow}$ if doing a dual lift.
Furthermore, for graphic (cographic) $h$, let $g$ be the graphic primal (dual) parity-check matrix from \cref{def:graphic cographic}. 

We start by defining the \emph{worm distribution} on the vector space given by subsets of $E$. 
To this end, let $w(S) = (p_{\uparrow / \downarrow}/(1-p_{\uparrow / \downarrow}))^{|S|}$ for $S \subset E$. 
Define the even-subgraph space $\Omega_0 = \ker(g^T)$, and the worm-space 
\begin{align*}
   \Omega_2 & \coloneqq \bigcup_{u,v \in [m]} \Omega(u,v), \quad \text{with }  
   \Omega(u,v) = \ker((g^T)_{[m] \setminus \{u,v\}}) \setminus \Omega_0 
\end{align*}
the configurations in which the two vertices $u,v$ have odd degree.
Compared to the even covers ($\ker(h^T)$ in the primal case, and $\ker((h^\perp)^T)$ in the dual case) the full space $\Omega_w \coloneqq \Omega_0 \cup \Omega_2$ is now enlarged by the even subgraphs not present in the even covers, and the worm configurations of all even subgraphs. 
The worm distribution $\omega_g$ on $S \subset E$ is defined as
\begin{align*}
  \omega_{g}(S) \propto w(S)\id(S \in \Omega_0) + \frac 1 {\binom {m} 2} w(S) \id(S \in \Omega_2),
\end{align*}
and penalizes those near-even-subgraph configurations such that the probability weight on $\Omega_0$ and on $\Omega_2$ is roughly the same.

To lift configurations sampled from the worm distribution to RC configurations, consider now the following lift, which is modified compared to the lift in \cref{lem:even covers rc coupling} only in that we start from configurations sampled from the worm distribution: 
Sample $S \leftarrow \omega_{g}$. 
If performing a primal lift, add each edge $e \in E \setminus A$ iid.\ with probability $p_{\uparrow}/(1-p_{\uparrow})$ to obtain $B$. 
Let the resulting marginal distribution on $B$ be $\phi_{\uparrow}$. Similarly, if performing a dual lift, add each edge $e \in E \setminus A$ iid.\ with probability $p_{\downarrow}/(1-p_{\downarrow})$, to obtain $B'$. Let the resulting marginal distribution on $E \setminus B'$ be $\phi_{\downarrow}$. Then $\phi_{\uparrow}$ and $\phi_{\downarrow}$ are both ``close to'' the RC model $\phi$ in the following way.

\begin{lemma}
\label{lem:subspace_loss}
 If $h$ is $\Delta$-graphic or $\Delta$-cographic, then
\begin{align*}
\frac{\phi_{\uparrow}(B)}{\phi(B)} \leq 2^{\Delta+1} \quad \text{ and} \quad\frac{\phi_{\downarrow}(B)}{\phi(B)} \leq 2^{\Delta+1},
\end{align*}
respectively, for any $B$.

\end{lemma}
We prove \cref{lem:subspace_loss} in \cref{sec:coupling_to_syndromes_and_even_covers}.

The first step in our proof of rapid mixing for the RC model is to show the existence of a good set of canonical paths for the even cover model \emph{through} the worm space $\Omega_w$. 
To this end, let us denote the even-cover model from the primal coupling as $\xi_{\uparrow} = \xi_{h,p/2}$ and the even-cover model from the dual coupling as $\xi_{\downarrow} = \xi_{h^\perp,(1-p)/(2-p)}$. 
We further define the primal and dual even-cover spaces as $\Omega_\uparrow = \ker(h^T)$ and $\Omega_\downarrow = \col(h)$.

\begin{definition}
A set of canonical paths $\Gamma$ for the even-cover model $\xi_{\uparrow/\downarrow}$ through the worm space $\Omega_w$ is one such that for all $\gamma \in \Gamma$, $\gamma_0,\gamma_{L(\gamma)} \in \Omega_{\uparrow/\downarrow}$, and $\gamma_i \in \Omega_w$ for all $0 < i < L(\gamma)$.
\end{definition} 

To construct such paths, we extend the constructions of~\cite{jerrum_polynomial-time_1993,guo_random_2017}. We show that these in fact give us a good set of canonical paths for the even-cover model in the following sense.
\begin{lemma}[SC dynamics for the worm model mixes rapidly] \label{lemma:worm_paths}
 Suppose that $h$ is $\Delta$-graphic or $\Delta$-cographic and coupled to $g \in \bin^{c \times m}$. Then there is a set of canonical paths $\Gamma$ with flow function $f$ for the even-cover model $\xi_{\uparrow / \downarrow}$ through $\Omega_w$ that satisfies  
 \begin{align*}
   \sum_{\gamma \ni (W,W') } f(\gamma) \leq 2^{\Delta+1} m^4 \omega_{g}(W)  
 \end{align*}
 
for any $W' = W \oplus e$. In the special case that $W'=W \cup e$, then
\begin{align*}
   \sum_{\gamma \ni (W,W') } f(\gamma) \le 2^{\Delta+1} m^4 \omega_{g}(W) \left(\frac{p_{\uparrow / \downarrow}}{1-p_{\uparrow / \downarrow}} \right) .
 \end{align*}
\end{lemma}

We prove \cref{lemma:worm_paths} in \cref{section:flows}.

Using \cref{lem:subspace_loss}, we can now lift these canonical paths to a good flow for the RC model for $\Delta$-graphic and $\Delta$-cographic $h$ to prove \cref{thm:main theorem}.

\begin{proof}[Proof of \cref{thm:main theorem}]

Suppose that $h$ is $\Delta$-graphic ($\Delta$-cographic). We show how to construct a good flow for $\phi_p$ using the canonical paths $\Gamma$ from \cref{lemma:worm_paths}.
Let $\gamma_{W_0,W_{\ell}}  = (W_0, \ldots, W_\ell) \in \Gamma$ with $\ell = L(\gamma)$ be the canonical path from $W_0$ to $W_\ell$ through $\Omega_w$. 
We will construct a flow, i.e., a distribution over paths $(Z_0, \ldots Z_{\ell'})$ for the RC model based on $\gamma$.

\paragraph{Construction of the flow}
Given $W_0 \in \Omega_{\uparrow/\downarrow}$, we construct $Z_0 \subset E$ using the coupling of \cref{lem:even covers rc coupling}. That is, we add every $e \in E\setminus W_0$ with probability $(p_{\uparrow / \downarrow})/(1-p_{\uparrow / \downarrow})$ to $W_0$,  and take the complement if taking a dual lift, to obtain $Z_0$.
In other words, for $W \subset Z$, let 
\begin{align*}
\delta_\uparrow(W,Z)=\left(\frac{p_{\uparrow}}{1-p_{\uparrow}}\right)^{|Z \setminus W|} \left(1-\frac{p_{\uparrow}}{1-p_{\uparrow}}\right)^{|E \setminus Z|}
\end{align*}
 and for $Z \subset W^c \coloneqq E \setminus W$,
\begin{align*}
\delta_\downarrow(W,Z)=\left(\frac{p_{\downarrow}}{1-p_{\downarrow}}\right)^{|Z^c \setminus W|} \left(1-\frac{p_{\downarrow}}{1-p_{\downarrow}}\right)^{|E \setminus Z^c|}.
\end{align*}
Then $\Pr[Z_0 = Z] = \delta_\uparrow(W_0,Z)$ for any $Z \supset W_0$ if doing a primal lift and $\Pr[Z_0 = Z] = \delta_\downarrow(W_0,Z)$ if doing a dual lift.

We now ``follow'' the underlying path in $\gamma$ as follows: 
\begin{itemize}
  \item If $W_{k+1} = W_k$, let $Z_{k+1} = Z_k$. 
  \item If $W_{k+1} = W_k \cup e$ for $e \notin W_k$,    
    \begin{itemize}
        \item If primal: let $Z_{k+1} = Z_k \cup e$.
        \item If dual: let $Z_{k+1} = Z_k \setminus e$
    \end{itemize} 
  \item If $W_{k+1} = W_k \setminus e$ for $e \in W_k$, 
    \begin{itemize}
        \item If primal: resample the edge, i.e., let $Z_{k+1} = Z_k$ with probability $\frac{p_{\uparrow}}{1-p_{\uparrow}}$ and $Z_{k+1} = Z_k \setminus e$ otherwise.
        \item If dual: resample the edge, i.e., let $Z_{k+1} = Z_k$ with probability $\frac{p_{\downarrow}}{1-p_{\downarrow}}$ and $Z_{k+1} = Z_k \cup e$ otherwise.
    \end{itemize}

\end{itemize}
One may check that this ensures that $\Pr(Z_k=Z|\gamma) = \delta_\uparrow(W_k,Z)$ for a primal lifting and $\Pr(Z_k=Z|\gamma) = \delta_\downarrow(W_k,Z)$ for a dual lifting.

To finish the construction of the flow, observe that after this procedure, at the end of $\gamma$, $Z_{L(\gamma)}$ remains correlated with $Z_0$. To remove the correlation, we re-randomize the edges $\{e_1, \ldots, e_k\} = E \setminus W_\ell$ not in $W_\ell$ where $k = |E \setminus W_\ell|$. Therefore let 
\begin{itemize}
  \item If primal: $Z_{\ell + i+1} = Z_{\ell + i } \setminus e_i$ with probability $1-\frac{p_{\uparrow}}{1-p_{\uparrow}}$ and $Z_{\ell + i +1} = Z_{\ell + i} \cup e_i$ otherwise. 
  \item If dual: $Z_{\ell + i+1} = Z_{\ell + i } \setminus e_i$ with probability $\frac{p_{\downarrow}}{1-p_{\downarrow}}$ and $Z_{\ell + i +1} = Z_{\ell + i} \cup e_i$ otherwise.
\end{itemize}
  
We therefore obtain a path $\lambda = (Z_0, Z_1, \ldots, Z_{\ell + k})$ to which we assign the weight 
\begin{align*}
  F(\lambda) = \sum_{\gamma \in \Gamma} f(\gamma) \Pr(Z = \lambda | \gamma)
\end{align*}
and observe that $L(\lambda) \le L(\gamma) + c$. 

Let $\Lambda(I,F) $ be the set of paths from $I$ to $F\subset E$. We check that $F$ is a valid flow for $\phi$.  
To this end for a path $\lambda = (Z_1, \ldots, Z_{L(\lambda)})$ let $Z_{-1} = Z_{L(\lambda)}$.
\begin{align*}
  \sum_{\lambda \in \Lambda(I,F) } F(\lambda) & = \sum_{\substack{U, V\\ U,V \in \Omega_{\uparrow/\downarrow} }} f(\gamma_{U,V}) \Pr(Z_0 = I, Z_{-1} =F|\gamma_{U,V})\\ 
  & = \sum_{\substack{U, V \\ U,V \in \Omega_{\uparrow/\downarrow} }}\xi_{\uparrow / \downarrow}(U)\xi_{\uparrow / \downarrow}(V) \delta_{\uparrow / \downarrow}(U,I)\delta_{\uparrow / \downarrow}(V,F)\\ 
  & = \phi(I) \phi(F), 
\end{align*}
where we have used that 
$\Pr(Z_0 = I, Z_{-1} =F|\gamma) = \delta_{\uparrow / \downarrow}(U,I)\delta_{\uparrow / \downarrow}(V,F)$, that $f$ is a flow for $\xi_{\uparrow/\downarrow}$, 
and that 
\begin{align*}
  \phi(S) = \sum_{W \in \Omega_{\uparrow/\downarrow}, S \subset E} \xi_{\uparrow/\downarrow}(W) \delta_{\uparrow / \downarrow}(W,S).
\end{align*}

\paragraph{Bounding the flow through a transition}
We now bound the flow through any edge $(Z,Z')$ in the primal lift. The proof for the dual lift will follow analogously. First, let us define $i(\gamma, W)$ to be the index of state $W$ in path $\gamma$, and let $k(W,e)$ be the index of edge $e$ in $E \setminus W_\ell$. Also define $r(p) = \frac{p}{1-p}$.
We need to separately consider the three cases (1) $Z' = Z \cup e$, (2) $Z' = Z \setminus e$, and (3) $Z' = Z $. 

\noindent \textbf{Case 1} If $Z' = Z \cup e$,  
\begin{align*}
  \sum_{\lambda \ni (Z,Z') } F(\lambda) &= \sum_{W \subset Z} \Bigg( \sum_{\gamma \ni (W, W \cup e)} f(\gamma) \Pr(Z_{i(\gamma, W)}=Z,Z_{i(\gamma, W)+1}=Z' |\gamma) \\ 
  & \qquad + \sum_{\gamma: \gamma_{-1}=W} f(\gamma) \Pr(Z_{\ell+k(W,e)}=Z,Z_{\ell+k(W,e)+1}=Z' |\gamma) \Bigg)\\
  &=\sum_{W \subset Z} \left( \sum_{\gamma \ni (W, W \cup e)} f(\gamma) \delta_{\uparrow }(W,Z) + \sum_{\gamma: \gamma_{-1}=W} f(\gamma) \delta_{\uparrow }(W,Z) r(p_{\uparrow }) \right)\\ 
  & = \sum_{W \subset Z } \delta_{\uparrow }(W,Z) \left( \sum_{\gamma \ni (W, W \cup e)} f(\gamma)  + \sum_{\gamma: \gamma_{-1}=W} f(\gamma)  r(p_{\uparrow }) \right) \\ 
  & \leq \sum_{W \subset Z } \delta_{\uparrow }(W,Z) \left( 2^{\Delta+1} n^4 \omega_{g}(W) r(p_{\uparrow }) + \xi_\uparrow(W) r(p_{\uparrow }) \right) \\
  & =  \left[2^{\Delta+1} m^4 \phi_\uparrow(Z) + \phi(Z)\right]r(p_{\uparrow })\\
  & \leq (2^{2\Delta+2}m^4  +1)r(p_{\uparrow })\phi(Z) \\
  &\leq  2^{2\Delta+3}m^4 \phi(Z) r(p_{\uparrow }),
\end{align*}
where we used \cref{lemma:worm_paths} for the first inequality and \cref{lem:subspace_loss} for the second inequality, and observed that $\phi_{\uparrow}(Z) = \sum_{W \subset Z} \omega_g(W) \delta_{\uparrow}(W,Z)$ and $\phi(Z) = \sum_{W \subset Z} \xi_\uparrow(W) \delta_{\uparrow}(W,Z)$ by \cref{lem:even covers rc coupling}.
Using the same arguments, we can bound the flow for cases 2 and 3.

\noindent \textbf{Case 2}
If $Z' = Z \setminus e$,
\begin{align*}
  \sum_{\lambda \ni (Z,Z') } F(\lambda) &= \sum_{W \subset Z} \Bigg( \sum_{\gamma \ni (W, W \cup e)} f(\gamma) \Pr(Z_{i(\gamma, W)}=Z,Z_{i(\gamma, W)+1}=Z' |\gamma) \\ 
  & \qquad + \sum_{\gamma: \gamma_{-1}=W} f(\gamma) \Pr(Z_{\ell+k(W,e)}=Z,Z_{\ell+k(W,e)+1}=Z' |\gamma) \Bigg)\\
  &=\sum_{W \subset Z} \left( \sum_{\gamma \ni (W, W \setminus e)} f(\gamma) \delta_{\uparrow}(W,Z)(1-r(p_{\uparrow})) + \sum_{\gamma: \gamma_{-1}=W} f(\gamma) \delta_{\uparrow}(W,Z) (1-r(p_{\uparrow})) \right)\\ 
  & = \sum_{W \subset Z} \delta_{\uparrow}(W,Z) \left( \sum_{\gamma \ni (W, W \setminus e)} f(\gamma)(1-r(p_{\uparrow}))  + \sum_{\gamma: \gamma_{-1}=W} f(\gamma)  (1-r(p_{\uparrow})) \right) \\ 
  & \leq \sum_{W \subset Z} \delta_{\uparrow}(W,Z) \left( 2^{\Delta+1} m^4 \omega_{g}(W)(1-r(p_{\uparrow})) + \xi_\uparrow(W) (1-r(p_{\uparrow})) \right) \\
  & = 2^{\Delta+1}m^4 (1-r(p_{\uparrow})) \phi_\uparrow(Z) + (1-r(p_{\uparrow})) \phi(Z) \\
  &\leq  2^{2\Delta+3}m^4 \phi(Z) (1-r(p_{\uparrow}))
\end{align*}

\noindent \textbf{Case 3}
If $Z' = Z$,
\begin{align*}
  \sum_{\lambda \ni (Z,Z') } F(\lambda) &= \sum_{W \subset Z} \Bigg( \sum_{\gamma \ni W} f(\gamma) \Pr(Z_{i(\gamma, W)}=Z,Z_{i(\gamma, W)+1}=Z |\gamma)  \\ 
  & \qquad + \sum_{\gamma: \gamma_{-1}=W} \sum_{i=1}^{|E\setminus W|}f(\gamma) \Pr(Z_{\ell+i}=Z,Z_{\ell+i+1}=Z |\gamma) \Bigg)\\
  &\leq \sum_{W \subset Z} \left( \sum_{\gamma \ni W} f(\gamma) \delta_{\uparrow}(W,Z) + c \sum_{\gamma: \gamma_{-1}=W} f(\gamma) \delta_{\uparrow}(W,Z) \right)\\ 
  & = \sum_{W \subset Z} \delta_{\uparrow}(W,Z) \left( \sum_{\gamma \ni W} f(\gamma)  + c\sum_{\gamma: \gamma_{-1}=W} f(\gamma)   \right) \\ 
  & \leq \sum_{W \subset Z} \delta_{\uparrow}(W,Z) \left( 2^{\Delta+1} m^4 \omega_{g}(W) + \xi_\uparrow(W) c \right) \\
  & =2^{\Delta+1}m^4 \phi_\uparrow(Z) + c \phi(Z) \\
  &\leq  2^{2\Delta+3}m^4 \phi(Z) c 
\end{align*}

\paragraph{Bounding congestion} To bound the congestion, we again go through the three cases. 

\noindent  \textbf{Case 1}: 
If $Z' = Z \cup e$, 
\begin{align*}
    \frac{1}{\phi(Z) P(Z,Z')} \sum_{\gamma, (Z,Z') \in \gamma} F(\gamma) L(\gamma) &\le \frac{c}{\phi(Z) P(Z,Z')} 2^{2\Delta+3}m^4 \phi(Z) r(p_{\uparrow}) \\
    &\le \frac{c^2}{\phi(Z) \min(1,\frac{p}{2(1-p)})} 2^{2\Delta+4}m^4 \phi(Z) r(p_{\uparrow}) \\
    &\le c^2 2^{2\Delta+5}m^4
\end{align*}
    
\noindent  \textbf{Case 2}: 
If $Z' = Z \setminus e$, 
\begin{align*}
    \frac{1}{\phi(Z) P(Z,Z')} \sum_{\gamma, (Z,Z') \in \gamma} F(\gamma) L(\gamma) &\le \frac{c}{\phi(Z) P(Z,Z')} 2^{2\Delta+3}m^4 \phi(Z) (1-r(p_{\uparrow})) \\
    &\le \frac{c^2}{\phi(Z) \min(1,\frac{1-p}{p})} 2^{2\Delta+4}m^4 \phi(Z) (1-r(p_{\uparrow})) \\
    &\le c^2 2^{2\Delta+5}m^4
\end{align*}

\noindent  \textbf{Case 3}: 
If $Z' = Z$, 
\begin{align*}
    \frac{1}{\phi(Z) P(Z,Z')} \sum_{\gamma, (Z,Z') \in \gamma} F(\gamma) L(\gamma) &\le \frac{c}{\phi(Z) P(Z,Z')} 2^{2\Delta+3}m^4 \phi_p(Z) c \\
    &\le c^2 2^{2\Delta+4}m^4
\end{align*}

Altogether, the congestion over any edge $(Z,Z')$ is therefore upper-bounded by 
\begin{align*}
  \rho(F) \le c^2 2^{2\Delta+5}m^4 =  O(c^2 2^{2\Delta+5}n^4),
\end{align*}
since by \cref{def:graphic cographic} $m = O(n)$

The proof for dual lifting congestion follows analogously, except with $p_{\downarrow}$ in place of $p_{\uparrow}$ and the analyses for $Z' = Z \cup e$ and $Z'=Z \setminus e$ flipped.
\end{proof}

\subsection{Slow mixing of the Code Swendsen-Wang dynamics}
\label{sub:slow mixing}

In this section, we show that there are classical codes and choices of the inverse temperature $
\beta$ for which the Code Swendsen-Wang dynamics faces a bottleneck and takes an exponentially long time to mix. 
Our arguments follow the general ideas of \textcite{gore_swendsenwang_1999}.

We consider the ferromagnetic \emph{3-spin Curie-Weiss model}. 
This model is defined by a parity check matrix $h$, whose rows are given by all strings of Hamming weight $3$, so the checks in $h$ define the complete $3$-ary hypergraph. This is a generalization of the Curie-Weiss model. Because of the permutation symmetry of the model, the energy only depends on the Hamming weight $|x|$ of a configuration $x$, and is given by 
\begin{align}
\label{eq:curie weiss}
  H(x)\equiv H(|x|) &= 2 \binom {|x|}{3} +2 \binom{|x|}{1}\binom{n-|x|}{2}
\end{align}
 with the associated Gibbs distribution $\pi(x) \equiv \pi(an) = Z^{-1}  \exp( - \beta H(an) ) $ where $a  = |x|/n$. 
The equilibrium probability of being in a configuration with Hamming weight $an$ is therefore given by $\sigma(a) \coloneqq N(an) \pi(an)$ with $N(an) = \binom{n }{an}$, which equals \cite[][Sec.~9]{feller_introduction_2009}
\begin{align}
  N(an) = \frac{1}{\sqrt {2 n a (1-a)}} \exp(n S(a) + \Theta(1/n)),
\end{align}
for $a \in \Omega(1)$
where $S(a) = -a \ln (a) - (1-a) \ln (1-a)$ is the entropy function. Here, the $\Theta(1/n)$ term captures corrections to Stirling's approximation \cite{feller_introduction_2009} as well as from the binomial coefficients in \cref{eq:curie weiss}. 
Altogether, we find 
\begin{align}
\label{eq:configuration prob}
  \sigma(a) = \frac{1}{Z \sqrt {2 n a (1-a)}} \exp(f(a)n + \Theta(1/n)),
\end{align}
where, setting $c = \beta n^2$, 
\begin{align}
  f(a) &= S(a) - c  \left(a^3/3 + a(1-a)^2\right)
\end{align}

\begin{lemma}
  There exists a critical $c^* > 0 $, such that the function $f$ has exactly two distinct global maxima at $a_0, a_1  \in (0,1)$. 
  These values satisfy $4.0324 < c^* < 4.0326$, $a_0 <  0.026$ and $a_1 =1/2$. 
\end{lemma}
\begin{proof}
Computing 
\begin{align}
  f'(a) = \ln(\frac{1-a}a) - c (2a - 1)^2,
\end{align}
we find that $f(a)$ has a local extremal point at $a=1/2 $ for all values of $c$. 
Moreover, the value of $f$ at this point is 
\begin{align}
  f(1/2) = S(1/2) - \frac{2c}{12} =  \ln 2 - \frac {2c}{12}.
\end{align}
Substituting the condition on $c$ at $f'(a) =0$ we find 
\begin{align}
  c(a) = \ln (\frac{1-a}a) /(2a - 1)^2,
\end{align}
which gives the equation
\begin{align}
   f(a) = S(a) - c(a)\left( a^3/3 + a(1-a)^2\right) = S(1/2) - \frac 1{6} c(a) = f(1/2).
 \end{align} 

This equation has three zeros are at $a = a_0, a_1, 1- a_0$ with $0.025 \le a_0 \le 0.026$ and $a_1 = 1/2$, but we can discard the last, since it corresponds to negative temperature. 
Moreover, we can evaluate
\begin{align}
  f''(a) = -\frac 1{a (1-a)} - 4c^*(2a-1) 
\end{align}
at the extremal points $a = a_0, a_1$ and find that they are the unique local and global maxima at the critical value $c^* =c(a_0) $ that satisfies $4.0324 < c^* < 4.0326$.
\end{proof}

Let $\mc B_0(\epsilon), \mc B_1(\epsilon) \subset \{0,1\}^n$ be balls of Hamming-radius $\epsilon$ around strings with Hamming weight $a_0$ and $a_1$, respectively, and $B_0(\epsilon), B_1(\epsilon) \subset \mb R$ be the respective balls on the real line. 
\begin{lemma}
The probability weight of $\pi$ is distributed as follows:
  \begin{enumerate}[label=\roman*.]
     \item $\pi(\mc B_0(\epsilon)) \ge \Omega(1/n)$
    \item $\pi(\mc B_1(\epsilon)) \ge \Omega(1/n)$
   \item $\pi(\mc B_0(\epsilon) \cup \pi(\mc B_1(\epsilon)) \ge 1 - \exp(-\Omega(n))$. 
   \end{enumerate} 
\end{lemma}
\begin{proof}
For part i.\ observe that for $|a-a_0| \in O(1/n)$, we have $|f(a) - f(a_0)| = O(1/n)$  and therefore $\sigma(a)/\sigma(a_0) \in O(1)$, which implies that $\int_{a \in B_0(\epsilon)}da \, \sigma(a)  \ge \Omega(1/n)$. 
Part ii.\ follows similarly. 

For part iii.\ observe that that for $\epsilon \in \Omega(1)$ and $a - a_0 \ge \epsilon$ we have  $0 \le f(a_0) - f(a) \in \Omega(1)$.
This implies an upper bound on the probability that a string $x$ has Hamming weight $a \notin B_0(\epsilon)\cup B_1(\epsilon)$ as
\begin{align}
  \sigma(a) \le \frac {\sigma(a)}{\sigma(a_0)}= \sqrt{\frac{a(1-a)}{a_0(1-a_0)}} \exp(-n(f(a_0)- f(a)) + \Theta(1/n))  = \exp(- \Omega(n)). 
\end{align}
Altogether we thus find
\begin{align}
  \pi(\mc B_0(\epsilon) \cup \pi(\mc B_1(\epsilon))= \int_{a \in [0,1]\setminus ( B_0(\epsilon) \cup B_1(\epsilon))} \sigma(a)= \exp(- \Omega(n)).
\end{align}
\end{proof}

\begin{theorem}
  The mixing time of the Code SW dynamics for the ferromagnetic 3-spin Curie-Weiss model \eqref{eq:curie weiss} at inverse temperature $\beta^* = c^*/n^2$ is $\exp(\Omega(n))$. 
\end{theorem}
\begin{proof}
Partition the configurations into $A = \mc B_0(\epsilon)$, $B = \mc B_1(\epsilon)$ and $C = \mc B_2(\epsilon) = \{0,1\}^n \setminus (\mc B_0(\epsilon) \cup \mc B_1(\epsilon))$ for $ \epsilon > 0 $. 

Consider starting from a configuration $x$ with roughly half of the checks satisfied at the critical inverse temperature $\beta^*$. 
This is a configuration in the disordered phase $\mc B_1(\epsilon)$. 
We are going to show that a Swendsen-Wang update of such a configuration is going to stay in the disordered phase with overwhelming probability. 

By Lemma~23, at the critical inverse temperature \(\beta^\ast\), both phases have at least inverse-polynomial stationary mass:
\[
\pi(A)\ge \Omega(1/n),
\qquad
\pi(B)\ge \Omega(1/n).
\]
We will show that the stationary probability flow out of \(B\) is exponentially small. We then use reversibility and a standard conductance lower bound to conclude that the mixing time is exponential.

We first show that the probability  flow  
\[
F(B \rightarrow A) \coloneqq \sum_{b \in B}\pi(b)\sum_{a \in A} P(b,a)
\]
 for the CSW transition probabilities $P(B,A)$ satisfies $F(B \rightarrow A) \le 2^{-\Omega(n)}$ (i), and then show that the flow $F(B \rightarrow C)\le 2^{-\Omega(n)}$ (ii).

We first show that
\[
F(B\to A)\le \exp(-\Omega(n)),
\tag{i}
\]
and then that
\[
F(B\to C)\le \exp(-\Omega(n)).
\tag{ii}
\]
Together these imply
\[
F(B\to B^c)
=
F(B\to A)+F(B\to C)
\le \exp(-\Omega(n)).
\]

(i) Let us begin by considering $x \in B$. This means that its Hamming weight $|x| = (1/2 + \delta)n$ with $|\delta| \le \epsilon $ exactly and the checks $E(x)$ that are satisfied by this configuration. We are going to form clusters starting from those checks according to step i.\ of the Code SW algorithm. 
Observe that the satisfied checks fall into two groups. 
One group is comprised of the checks $C_0(x) \subset E$ that are fully contained in the set $Z(x) = \{i \in [n]: x_i = 0\}$. 
The other group is comprised of the checks $C_1(x) \subset E$ which touch one variable in $Z(x)$ and two variables in $O(x) = [n] \setminus Z(x)$. 

Now, we will keep checks from $E(x)$ with probability $p = 1- e^{-2 \beta^*} = 1- e^{-2 c^*/n^2 }$. Consider $C_1(x)$. Restricted to $O(x)$, $C_1(x)$ induces the complete graph on $n' = (1/2 + \delta)n$ vertices with $n'$ copies of the same edge, since every check in $C_1(x)$ touches exactly two vertices in $O(x)$, and for every edge between two vertices in $O(x)$ there are $n'$ vertices in $Z(x)$. 
We will now consider the graph $G(n',p')$ induced on $O(x)$ after iid.\ subsampling $C_1(x)$ with probability $p$, yielding a set $S \subset C_1(x)$. This graph has an edge between vertices $i,j \in O(x)$ if at least one hyperedge contains both $i$ and $j$, which for each edge occurs  with probability 
\[
p' =  1- (1-p)^{n-n'} = p(n-n') - O(p^2n'^2) = \frac{c^*}{2n'} ( 1 - 4 \epsilon^2 ) + O(1/n^3) \le \frac{2.02}{n'}
\]
for sufficiently large $n$. 

We now argue that the resulting graph has an extensive number of isolated vertices except with exponentially small probability.
To this end we invoke the following result of \textcite{ghosh_concentration_2011}. 
\begin{lemma}[Concentration bound for isolated vertices \cite{ghosh_concentration_2011}]
\label{lem:concentration isolated}
  The number $N$ of isolated vertices in the random graph $G(n,p)$ satisfies the tail inequality 
  \begin{align}
    \Pr[N \le \mb E[N] -t ] \le \exp(- \frac{t^2}{4\mb E[N]}),    
  \end{align}
  where $\mb E[N]$ denotes the expectation of $N$.
\end{lemma}

We also have that the expected number of isolated vertices $N$ in $G(n',p')$ is given by \cite{bollobas_evolution_1984,ghosh_concentration_2011}
\begin{align}
  \mb E[N] = (n - n') (1-p')^{n'-1} \ge  (n-n')\left(1- \frac {2.02} {n'}\right)^{n'-1} \ge  0.065 n. 
\end{align}
where the inequalities hold for sufficiently large $n'$ and $\epsilon < 0.001$.

As a consequence of \cref{lem:concentration isolated}, the probability $\Pr[N\le 0.06n] = \exp(- \Omega(n))$ that the number of vertices in $O(x)$ that is isolated in the graph induced by the subsampled $S \subset C_1(x)$ is at least $0.06n$ except with inverse exponential probability. 

Since the corresponding bits of the updated string are just uniformly random after step ii.\ of the Code SW chain, the updated configuration $x'$ will have Hamming weight at least $0.029 n > a_0n$  except with probability $\exp(- \Omega(n))$ by the  Hoeffding large deviation bound for a sum of uniform random variables. 
 
 Altogether we therefore have 
 \begin{align}
   F(B \rightarrow A ) = \sum_{b\in B}\pi(b)P(b,A)
\le
\exp(-\Omega(n))\sum_{b\in B}\pi(b)
\le
\exp(-\Omega(n)).. 
 \end{align}

(ii) Detailed balance of the Code SW dynamics implies that the flow 
\begin{multline}
  F(B \rightarrow C) = \sum_{b \in B, c \in C} \pi(b) P(b,c) = \sum_{b \in B, c \in C} \pi(c) P(c,b) 
  \le \sum_{c \in C} \pi(c) \le \exp(- \Omega(n)). 
\end{multline}
from $B$ to $C$ is exponentially small.

Combining (i) and (ii),
\[
F(B\to B^c)\le \exp(-\Omega(n)).
\]
By reversibility, the same bound holds for \(F(B^c\to B)\). Let \(S\in\{B,B^c\}\) be the set with \(\pi(S)\le 1/2\). By Lemma~23,
\[
\pi(S)\ge \Omega(1/n),
\]
since \(\pi(B)\ge \Omega(1/n)\) and \(\pi(B^c)\ge \pi(A)\ge \Omega(1/n)\). Hence
\[
\Phi(S)=\frac{F(S\to S^c)}{\pi(S)}
\le \exp(-\Omega(n)).
\]
The standard conductance lower bound \cite[Chapter~7]{levin_markov_2009} therefore gives
\[
t_{\mathrm{mix}}=\exp(\Omega(n)).
\]

\end{proof}

\section*{Acknowledgements}

We are grateful to Yaodong Li for illuminating discussions on lattice gauge theories. 
DH was supported by a Simons postdoctoral fellowship through DOE QSA and NSF QLCI Grant No. 2016245, and from the Swiss National Science Foundation through Ambizione Grant No.\ 223764. NJ and UV were supported by NSF Grant CCF-231173, NSF QLCI
Grant 2016245 and DOE grant DE-SC0024124.

\printbibliography
\appendix
\section{Code Swendsen-Wang dynamics is faster than single-check dynamics}
\label{app:sw vs sc}

In this section, we prove \cref{lem:sw vs sc} by directly following the proof of \textcite{ullrich_swendsen-wang_2014}.

For convenience of the proof, we define the (lazy) \emph{single-check (SC) dynamics} of the random cluster model via the following update rule. 
\begin{enumerate}
  \item With probability $1/2$ let $B = A$. 
  \item Otherwise, choose a uniformly random check $e \in E$.
  \begin{itemize}
    \item If $k(A) = k(A \cup e)$, let $B = A \cup e$ with probability $p$, and $B = A \setminus e$ with probqability $1-p$. 
    \item If $k(A) \neq k(A \cup e)$, let $B = A \cup e$ with probability $p/2$, and $B = A$ with probability $1 - p/2$.
  \end{itemize}
  \item Output $B$.
\end{enumerate}
We use the standard equivalences between two Markov chains, and apply it to the SC update compared to the Metropolis update. 
To do this, we define the \emph{gap} of a Markov chain with transition matrix $P$ on state space $\Omega$ as
\begin{align*}
\Delta(P ) = 1 - \max\{|\lambda| : \lambda \text{ is an eigenvalue of } P, \lambda \ne 1 \}.
\end{align*}
The spectral gap is an equivalent characterization of  the mixing time as \cite[][Theorem 12.3 \& 12.4]{levin_markov_2009}, stated here, following \cite{ullrich_swendsen-wang_2014}
\begin{align*}
  \Delta(P)^{-1} - 1 \le \tau(P) \le 
  \log \left( \frac{2e}{\min_{S \in \Omega} p(S) } \right) \Delta(P)^{-1}
\end{align*}

\begin{theorem}[\cite{diaconis_comparison_1993}] 
Let $P$ and $Q$ be the transition matrices of two Markov chains over a state space $X$ with the same unique fixed point $\pi$, satisfying 
\begin{align*}
  P(x,y) \le Q(x,y) \le c P(x,y)
\end{align*}
for all $x\ne y \in X$ and some $c > 0$. Then 
\begin{align*}
  \Delta(P) \le \Delta (Q) \le c \Delta(P).   
\end{align*}
\end{theorem}
We now observe that 
\begin{align*}
  P_{\text{SC}}(A,B) \le P_{\text{Metropolis}}(A,B) \le 2 P_{\text{SC}}(A,B). 
\end{align*}
and subsequently follow the proof of \textcite{ullrich_swendsen-wang_2014} for the SC dynamcis. 
To this end, let us introduce some notation. 
Define the function space $L_2(\pi) \coloneqq(\mb R^\Omega, \pi)$ in which the inner product is given by 
\begin{align*}
   \langle f,g \rangle_\pi = \sum_{x \in \Omega} f(x)f(g) \pi(x),
\end{align*}
and $\norm{f}_\pi \coloneqq \langle f, f \rangle_\pi$. 
Define the operator $P$ as 
\begin{align*}
  P: & L_2(\pi) \rightarrow L_2(\pi)\\
  Pf(x) & \coloneqq \sum_{y \in \Omega} P(x,y) f(y). 
\end{align*}
Furthermore we have the operator norm of $P$, $\norm{P}_\pi \coloneqq \norm{P}_{L_2(\pi) \rightarrow L_2(\pi)} = \max_{\norm{f}_\pi \le 1 } \norm{Pf}_\pi$.

To compare SW and SC dynamics, we define two mappings. The first one, $M: \Omega \rightarrow X \times \Omega$ lifts a RC configuration to a FK configuration, the second one, $T_e$ updates the FK model. 
Let the set of configurations satisfying a set of checks $A \subset E$ be given by 
\begin{align*}
  X(A) &\coloneqq \ker (h_A).
\end{align*}
Then the two mappings are given by
\begin{align*}
 M(B, (x,A)) &\coloneqq 2^{-k(A)} \id(A = B) \id(x \in X(A)), \\
  T_e((x,A), (y,B)) &\coloneqq \id(x = y ) \begin{cases}
    p, & B = A \cup e \text{ and }  e \in E(x) \\
    1-p, & B = A \setminus e \text{ and }  e \in E(x) \\
    1, & B = A  \text{ and }  e \notin E(x) \\
    0, & \text{otherwise}.
  \end{cases}
\end{align*}
As above, these define operators $M :  L_2(\mu) \rightarrow L_2(\phi)$, and $T_e: L_2(\mu ) \rightarrow L_2(\mu)$. The adjoint of $M$ is given by 
$M^*((x,A),B) = \id(A = B)$, since
\begin{align*}
  \langle f, Mg \rangle_\phi&  = \sum_{B \in \Omega} f(B)  Mg(B) \phi(B)\\ 
   & \propto \sum_{B \in \Omega} f(B) \left(\sum_{(x,A) \in X \times \Omega} 2^{-k(A)} \id(A = B) \id(x \in X(A))  \right)\left(\frac p {1-p}\right)^{|B|} 2^{k(B)} \\ 
   & =\sum_{(x,A) \in X \times \Omega} \left(\sum_{B \in \Omega} \id(A = B) f(B)   \right)g( x,A) \left(\frac p {1-p}\right)^{|A|}\id(A \subset E(x) )  \\ 
   & = \sum_{(x,A) \in X \times \Omega} \left(\sum_{B \in \Omega} \id(A = B) f(B)\right)   g( x,A) \mu(x,A)  = \langle M^*f, g \rangle_\mu
\end{align*}.

\begin{lemma}
  Let $M$, $M^*$ and $T_e$ be the operators defined above. We have 
  \begin{enumerate}[label=\roman*.]
    \item $M^*M$ and $T_e$ are self-adjoint in $L_2(\mu)$. 
    \item $MM^*(A,B) = \id(A=B)$ and thus $M^*MM^*M = M^*M$. 
    \item $T_e T_e = T_e $ and $T_e T_{e'} = T_{e'} T_e $ for all $e, e' \in E$. 
    \item $\norm{T_e}_\mu = 1$ and $\norm{M^*M}_\mu = 1$. 
  \end{enumerate}
\end{lemma}
\begin{proof}
  (i) Self-adjointness of $M$: $\langle f, M^*M g \rangle_\mu = \langle M^*M f, g \rangle_\mu$. 

  To show self-adjointness of $T_e$, we can use that 
Hence, we can use that 
\begin{align*}
  \mu(x,B \setminus e)\id(e \in B) \id(e \in E(x)) &= \left( \frac{1-p}p\right) \mu(x,B) \id(e \in E(x)) \\ 
  \mu(x,B \cup e) \id(e \notin B)\id(e \in E(x)) &= \left( \frac p {1-p}\right) \mu(x,B)\id(e \in E(x)) 
\end{align*}
to find 
\begin{align*}
  & \sum_A f(x,A) p  g(x,A \cup e)\id(e \notin A)  \id(e \in E(x))\mu(x,A)  \\ 
  &  = \sum_B f(x,B \setminus e) p  g(x,B)\id(e \in B)\id(e \in E(x))\mu(x,B \setminus e) \\ 
  & = \sum_B f(x,B \setminus e) (1-p)  g(x,B)\mu(x,B ) ,
\end{align*}
and likewise 
\begin{align*}
    & \sum_A f(x,A)(1-p) g(x,A \setminus e)\id(e \in A) \id(e \in E(x))\mu(x,A)  \\ 
  &  = \sum_B f(x,B \cup e) (1-p)  g(x,B)\id(e \notin B)\id(e \in E(x))\mu(x,B \cup e) \\ 
  & = \sum_B f(x,B \cup e) p  g(x,B)\mu(x,B) \id(e \in E(x)) . 
\end{align*}

(iii) This follows from the fact that the transition probabilities depend only on the coordinates $x$ of a FK configuration, which are not changed by the update. 

(iv) Follows from (i-iii) since $\norm{T_e}_\mu = \norm{T_e^2}_\mu$ by (iii) and $\norm{T_e^2}_\mu = \norm{T_e}^2_\mu$ by self-adjointness of $T_e$. 

\end{proof}

Next, we express the Swendsen-Wang and single-check updates in terms of $M$ and $T_e$.

\begin{lemma}[\cite{ullrich_swendsen-wang_2014}] Let $T_e,M,M^*$ be the operators defined above. Then
  \begin{enumerate}[label=(\roman*)]
    \item $P_{\text{SW}} = M \left(\prod_{e \in E}T_e\right) M^*$
    \item $P_{\text{SC}} = \frac{I}{2} + \frac{1}{2 |E|}  M \left(\sum_{e \in E}  T_e\right) M^*$
  \end{enumerate}
\end{lemma}
\begin{proof}
  To see (i), we observe that $M$ and $T_e$ generate the update steps of the SW algorithm via left-multiplication. Given $B$ applying $M$ from the right chooses a uniformly random bit string $x$ compatible with $B$. Likewise, $T_e$ updates the cluster configuration by generating a new cluster configuration in which a check $e \in E(x)$ is kept/added to the cluster with probability $p$ or removed with probability $1-p$. 

  For $B \subset E(x)$, we have that 
  \begin{align*}
    \left(\prod_{e \in E}T_e\right) ((x,B),(x,A)) & =  \sum_C\underbrace{\left(\prod_{e \notin E(x)}T_e\right) ((x,B),(x,C))}_{\id(B=C)} \left(\prod_{e \notin E(x)}T_e\right) ((x,C),(x,A)) \\
    & = p^{|A|}(1-p)^{|E(x)|-|A|} \id(A \subset E(x))
  \end{align*}
  which gives, using that $M^*$ acts trivially,
  \begin{multline}
    \sum_{x} M(B,(x,B)) \left(\prod_{e \in E}T_e\right) ((x,B),(x,A)) \\= \sum_x\id (x \in X(A) \cap X(B)) p^{|A|}(1-p)^{|E(x)|-|A|} 2^{-k(B)} \equiv P_{SW}(B,A)
  \end{multline}

  To see (ii), we need to show that the transition depends on whether or not the endpoints of $e$ are connected through $A$. 

  For a pair $(x,B)$ with $x \in X(B)$ we have 
  \begin{align*}
    T_e((x,B),(x,A)) &= p \id(e \in E(x))( \id (A = B \cup e)) - \id(A = B\setminus e)) \\&\quad+ \id(e \notin E(x))\id(A=B) + \id(e \in E(x)) \id (A = B \setminus e)\\
    &  = p \id(e \in E(x))( \id (A = B \cup e)) - \id(A = B\setminus e)) + \id (A = B \setminus e)
  \end{align*}

Now, let us compute 
\begin{align*}
  M T_e M^* (A,B)&   = \sum_{x \in X(A)} 2^{-k(A)} T_e((x,B), (x,A))\\ 
  &  =  \id(B = A\setminus e) + p ( \id (A = B \cup e)) - \id(A = B\setminus e))\sum_{x \in X(A)} 2^{-k(A)} \id(e \in E(x)),
\end{align*}
where we find with $\id_e(A) =\id(\ker(h_A) = \ker(h_{A \cup e}))$
\begin{align*}
  \sum_{x \in X(A)} 2^{-k(A)} \id(e \in E(x)) = \id_e(A) + \frac 12 ( 1- \id_e(A))  = \frac 12 (1 + \id_e(A)). 
\end{align*}
This gives exactly the transition matrix $P_e$ from above. 
\end{proof}

\begin{proof}[Proof of \cref{lem:sw vs sc}]
The remainder of the proof follows from Sec.~6.1 of \textcite{ullrich_swendsen-wang_2014}. 
\end{proof}

\section{Coupling to syndromes and even covers}
\label{sec:coupling_to_syndromes_and_even_covers}
We will now show that the RC measure is coupled both to the syndrome/satisfied check Gibbs distribution and the even-cover model. We will need the following lemma.

\begin{lemma}
\label{lem:kernel colspace intersection}
    Let $h,g $ be matrices. 
    \begin{align*}
        \dim(\ker(A) \cap \col(B)) = \dim(\ker(AB)) - \dim(\ker(B))
    \end{align*}
\end{lemma}

\begin{proof}[Proof of \cref{lem:even covers rc coupling}]

\emph{(dual lift)}
Note that the distribution of syndromes $S$ is given by 
\begin{align*}
  \zeta(S) \propto e^{- 2 \beta |S|}\id(S \in \col(h)) = e^{- 2 \beta |S|}\id(S \in \ker(g^T)) \propto \xi_{g,(1-p)/(2-p)},  
\end{align*}
where 
$g$ is such that $\col(g)= \col(h)^\perp $.

This yields 
\begin{align}
\Pr(B) &= \sum_{S} \zeta(S)(1-p)^{|B\setminus S|} p ^{|E \setminus B |} \id (S \subset B) \nonumber\\
& = \sum_{S} (1-p)^{|S|}(1-p)^{|B\setminus S|} p ^{|E \setminus B |} \id(1_{S} \in \col(h) \wedge S \subset B ) \nonumber\\ 
& = \left( \frac {1-p} {p}\right )^{|B|} \sum_{S} \id(1_{S} \in (\col(h) \cap \ker(\id_{E \setminus B})) \label{eq:syndrome calculation 1}\\
& = \left( \frac p{1-p} \right )^{|E \setminus B|} 2^{\dim(\ker(h_{E \setminus B})) - \dim(\ker (h))} \propto \phi(E \setminus B), \label{eq:syndrome calculation 2}
\end{align}
where we have used that $S \subset B \Leftrightarrow 1_S \in \ker(\id_{E \setminus B})$, where $\id_{E \setminus B }$ is the projector onto the rows indexed by $E \setminus B$ in \cref{eq:syndrome calculation 1}, and \cref{lem:kernel colspace intersection} in \cref{eq:syndrome calculation 2}.

\emph{(primal lift)}
  We have 
  \begin{align}
    \Pr(B) & = \sum_S \xi(S) 
    \left( \frac p{1-p} \right)^{|B \setminus S|}
    \left(1 -  \frac p{1-p} \right)^{|E \setminus B|} \id( S \subset B \wedge S \text{ is even} ) \nonumber
    \\ 
    & = \sum_S 
    \left(\frac p {1-p}\right)^{|S|} \left( \frac p{1-p} \right)^{|B \setminus S|}\left(1 -  \frac p{1-p} \right)^{|E \setminus B|} \id( 1_S  \in \ker(\id_{E \setminus B}) \cap \ker(h^T) ) \nonumber \\
     & = \left(\frac p {1-2p}\right)^{|B|} \sum_S 
    \id( 1_S  \in \underbrace{\ker(\id_{E \setminus B})}_{= \col(\id_B)} \cap \ker(h^T) ) \nonumber \\ 
    & = \left(\frac p {1-2p}\right)^{|B|} 2^{\dim(\ker(h_B^T)) - \dim(\ker{\id_B})} \label{eq:ec-rc calculation 1}\\
    & = \left(\frac p {1-2p}\right)^{|B|} 2^{\dim(\ker(h_B)) + e - v- (e - |B|)}  
    \propto \phi_{2p}(B), \nonumber
  \end{align}
  where in \cref{eq:ec-rc calculation 1} we used \cref{lem:kernel colspace intersection} and in the last step we used the dimension formula
\begin{align*}
  \dim(\col(h)) = v - \dim(\ker(h)) = e - \dim(\ker(h^T)) = \dim(\col(h^T)).
\end{align*}
\end{proof}

\begin{proof}[Proof of \cref{lem:subspace_loss}]

We follow the same argument as in the proof of \cref{lem:even covers rc coupling}.

\emph{(dual lift)}
Let $H = (h|c_1, \ldots, c_\Delta) \in \bin^{c \times (n + \Delta)}$  be a column generator matrix of $\ker(g^T)$ where we have added linearly-independent columns $c_1, \ldots,c_\ell \notin \col(h)$ to $h$, and let $p' = (1-p)/(2-p)$.
We follow a calculation analogous to \cite[Lemma 3.1]{guo_random_2017}
 \begin{align}
\phi_\uparrow(E \setminus B) &= \sum_{S} \omega_{g,p'}(S)(1-p)^{|B\setminus S|} p ^{|E \setminus B |} \id (S \subset B) \nonumber \\
& \propto \sum_{S} (1-p)^{|S|}(1-p)^{|B\setminus S|} p ^{|E \setminus B |} \id( S \subset B )(\id(S \in \Omega_0) + \frac{1}{n^2} \id(S \in \Omega_2)) \nonumber \\
& \propto \left( \frac p{1-p} \right )^{|E \setminus B|} \sum_{S} \id( S \subset B )(\id(S \in \Omega_0) + \frac{1}{n^2} \id(S \in \Omega_2)) \nonumber \\ 
& =  \left( \frac p{1-p} \right )^{|E \setminus B|} \sum_{S}
\id(1_{S} \in (\underbrace{\ker(g^T)}_{=\col(H)} \cap \ker(\id_{E \setminus B})) \nonumber\\ & \qquad + \frac 1 {n^2}\sum_{i \ne  j \in [n+\Delta]}  \id(1_{S} \in (\underbrace{\ker((g^T)_{V(i,j)})}_{=\col(H(V(i,j)))})) \cap \ker(\id_{E \setminus B})) \nonumber \\
& =  \left( \frac p{1-p} \right )^{|E \setminus B|} \bigg(2^{\dim(\ker(H_{E \setminus B}))- \dim(\ker(H))} \nonumber\\ 
& \qquad + \frac 1 {n^2}\sum_{i \ne  j \in [n+\Delta]} 2^{\dim(\ker(H(V_{i,j})_{E \setminus B})) - \dim(\ker(H(V_{i,j})))}\bigg) \nonumber\\ 
& \le  \left( \frac p{1-p} \right )^{|E \setminus B|} 2^{- \dim(\ker(h))} 2^\Delta 2^{\dim(\ker(h_{E \setminus B}))} 
\left( 1 + \binom{n}{2}/n^2 \right) \label{eq:explain coupling}\\
& \propto \frac 32 2^\Delta \phi(E \setminus B) \nonumber
\end{align}
where we have defined $V(i,j) = [n+\Delta] \setminus \{i,j\}$. 
In line \eqref{eq:explain coupling}, we used that $\dim(\col(H_A)) \ge \dim(\col(h_A))$ for $A \subset E$ and therefore, by the dimension formula,  $\dim(\ker(H_A)) \le \dim(\ker(h_A)) + \Delta$. 
Moreover, $\dim(\ker(H(V(i,j))_A))\le \ker(H_A)$.
Similarly, since 
$\dim(\col(H)) = \dim(\col(h))+ \Delta$, we have $\dim(\ker(H)) = \dim(\ker(h))$. 
This yields the claim for the dual lift.

For the primal lift, we follow the same reasoning.  
\end{proof}

\section{Canonical paths for the even subgraph model} \label{section:flows}

In this section, we construct the flows for \cref{lemma:worm_paths}. 
These flows are the \emph{unwinding} flows by Jerrum and Sinclair and the proof follows a standard canonical paths argument~\cite{jerrum_polynomial-time_1993}.
\begin{proof}[Proof of \cref{lemma:worm_paths}]
    We give the proof for the dual case. The primal follows analogously.

    We construct the canonical paths as follows.
    For any pair of states $A,B \in \col(h)$, which we can interpret as two even subgraphs in the graph described by $g$ due to the $\Delta$-graphic property, we will construct a path from $A$ to $B$ through the state space $\Omega_w$. 
    Consider the symmetric difference $A \oplus B$, which is also an even subgraph in $g$. Place a canonical ordering on cycles in $g$ and a canonical ordering of edges within each cycle, so that $A \oplus B$ is a disjoint union of cycles, ordered by this canonical ordering. Use the ordering of edges and cycles to decompose $A \oplus B$ into a sequence of edges $(e_1, \ldots, e_{\ell})$ for $\ell \le c$. Then traverse from $A$ to $B$ by taking the path determined by this sequence of edges $\gamma_{A,B} = (A, A \oplus e_1, A \oplus e_1 \oplus e_2, \ldots, A \oplus e_1 \oplus \ldots e_{\ell-1}, B)$. All states along this path are in $\Omega_w$ because at most two vertices have odd degree at all times. Assign this path a weight of $f(\gamma_{AB})=\xi(A) \xi(B)$.

    Now we bound the flow through any transition from $W$ to $W' = W \oplus e$. For configurations $A,B \in \col(h)$, let $\Phi(A,B)=W\oplus A \oplus B$. This is an injective map because given $(W,W')$ and $U=\Phi(A,B)$, we can recover $A$ and $B$ in the following way: Since $U \oplus W = A \oplus B$, then there is a canonical ordering on the edges in $U \oplus W$, including $e$. For any edge before $e$, its status is that in $B$ and for any edge after $e$, its status is that in $A$. Finally, because $U \oplus W$ gives you the symmetric difference between $A$ and $B$, then one can infer the remaining parts of $A$ and $B$.

    Recall that $w(A)=p_\downarrow^{|A|}(1-p_\downarrow)^{|E\setminus A|}$ since we are in the dual case. 
    Define $Z_\downarrow = \sum_{A\in \col(h)}w(A)$, $Z_0 = \sum_{A\in \Omega_0}w(A)$, and $Z_2 = \sum_{A\in \Omega_2}w(A)$. 
    
    We first bound the ratios between these quantities. First define the unnormalized vector $\ket{T}=(\ket{0} + \frac{p_\downarrow}{1-p_\downarrow}\ket{1})^{\otimes m}$ and $S = \sum_{A \in \col(h)} \ket{A}$. Note that $p_\downarrow \le 1/2$, so $H \ket{T}$ (the $m$-qubit Hadamard transform) only has positive coefficients, and similarly $H \ket{S}$. Then $Z_{\downarrow} = \braket{T}{S}=  \bra{T} H H \ket{S}$ is a summation of only positive elements. Letting $X_a$ be the $X$-Pauli tensor product described by $a$, then $\bra{T} X_a\ket{S} = \bra{T} HH X_a \ket{S} = \bra{T} H Z_aH\ket{S} \leq Z_\downarrow$ because this is now the same sum but with potential minus signs. Then letting $D$ be all $2^{\Delta}$ affine shifts for $\col(h) \subset \ker(g)$ to cover $\ker(g)$, then $Z_0 = \sum_{a \in D} \bra{T} X_a \ket{S} \leq \sum_{a \in D} Z_{\downarrow} \leq 2^{\Delta} Z_{\downarrow}$. 
    Similarly, \textcite{guo_random_2017} showed that $Z_2 \leq \binom{m}{2}Z_0$ in their Lemma 8.
    
    This allows us to bound the flow through the transition $(W,W')$ as 
    \begin{align*}
        \sum_{\gamma \ni (W,W')} f(\gamma) &= \sum_{\substack{A,B \in \col(h): \\ \gamma_{AB} \ni (W,W')} } \xi_\downarrow(A) \xi_\downarrow(B) \\
        &= \sum_{\substack{A,B \in \col(h): \\ \gamma_{AB} \ni (W,W')} } \frac{w(A)w(B)}{Z_\downarrow^2} \\
        &= \sum_{\substack{A,B \in \col(h): \\ \gamma_{AB} \ni (W,W')} } \frac{w(W)w(\Phi(A,B))}{Z_\downarrow^2} \\
        &\le w(W) \sum_{U \in \Omega_w} \frac{w(U)}{Z_\downarrow^2} \\
        &= w(W) \frac{Z_0 + Z_2}{Z_\downarrow^2} ,
    \end{align*}
    where the third equality uses that in every index in which $A$ and $B$ agree, $W$ and $\Phi(A,B)$ agree with $A$ and $B$, and the inequality used the fact that $\Phi$ is an injection.

    Continuing,
    \begin{align*}
        \sum_{\gamma \ni (W,W')} f(\gamma) &\le \frac{w(W)}{Z_\downarrow} \frac{Z_0 + Z_2}{Z_\downarrow} \\
        &\le \left( \binom{m}{2} \omega(W) \right) \left( 2^{\Delta} + \binom{m}{2} 2^{\Delta} \right) \\
        &\le 2^{\Delta + 1} m^4 \omega_g(W).
    \end{align*}

    In the special case that $W' = W \cup e$, then let $\Phi(A,B)=W' \oplus A \oplus B$ instead. Following the same proof, we see
    \begin{align*}
        \sum_{\gamma \ni (W,W')} f(\gamma) &\leq w(W') \frac{Z_0 + Z_2}{Z_\downarrow^2} \\
        &\leq w(W) \frac{Z_0 + Z_2}{Z_\downarrow^2} \frac{p_\downarrow}{1-p_\downarrow} \\
        &\le 2^{\Delta + 1} m^4 \omega_g(W) \frac{p_\downarrow}{1-p_\downarrow}
    \end{align*}
\end{proof}

\end{document}

%% file: mymath.tex
\newclass{\sh}{\#}
\newclass{\bis}{BIS}

\newclass{\stoqma}{StoqMA}
\newclass{\classP}{P}
\newclass{\bqp}{BQP}
\newclass{\qcam}{QCAM}
\newclass{\postbqp}{postBQP}
\newclass{\posta}{postA}
\newclass{\postiqp}{postIQP}
\newclass{\classa}{A}
\newclass{\bpp}{BPP}
\newclass{\fbpp}{FBPP}
\newclass{\pp}{PP}
\newclass{\cocp}{coC_=P}
\newclass{\ph}{PH}
\newclass{\np}{NP}
\newclass{\conp}{coNP}
\newclass{\gapp}{GapP}
\newclass{\approxclass}{Apx}
\newclass{\gapclass}{Gap}
\newclass{\sharpP}{\#P}
\newclass{\ma}{MA}
\newclass{\am}{AM}
\newclass{\qma}{QMA}

\newclass{\maxcut}{MAXCUT}
\newclass{\sat}{SAT}
\newclass{\maxtwosat}{MAX2SAT}
\newclass{\twosat}{2SAT}
\newclass{\threesat}{3SAT}
\newclass{\sharpsat}{\#SAT}
\newclass{\classx}{X}

\newtheorem{theorem}{Theorem}

\newtheorem{definition}[theorem]{Definition}

\newtheorem{lemma}[theorem]{Lemma}

\newtheorem{claim}[theorem]{Claim}

\newtheorem{corollary}[theorem]{Corollary}

\DeclareMathOperator{\supp}{supp}

\DeclareMathOperator{\row}{row}
\DeclareMathOperator{\col}{col}

\newcommand{\mc}{\mathcal}
\newcommand{\mb}{\mathbb}

\newcommand{\id}{\mathds{1}}

\DeclareMathOperator*{\Eb}{\mb E}

\newcommand{\bin}{\{0,1\}}
\newcommand{\pmbin}{\{-1,+1\}}

\newcommand{\proj}[1]{\ket{#1}\bra{#1}}